\documentclass[aps,prb,showpacs,notitlepage,floatfix, superscriptaddress,
 twocolumn]{revtex4-1}
\usepackage{bbm}
\usepackage{mathrsfs}
\usepackage{epsfig}
\usepackage{graphicx}
\usepackage{amsfonts}
\usepackage[figuresright]{rotating}
\usepackage{amssymb}
\usepackage{amsmath}
\usepackage{dcolumn}
\usepackage{bm}
\usepackage{braket}
\usepackage[colorlinks,linkcolor=blue,anchorcolor=blue,citecolor=blue,urlcolor=blue]{hyperref}
\usepackage{multirow}
\usepackage{longtable}
\usepackage{pbox}
\usepackage{float}
\usepackage[normalem]{ulem}

\def\be{\begin{equation}} \def\ee{\end{equation}}
\def\bp{\begin{pmatrix}} \def\ep{\end{pmatrix}}
\def\bea{\begin{eqnarray}}
\def\eea{\end{eqnarray}}
\def\beaa{\begin{equation}\begin{aligned}}
\def\eeaa{\end{aligned}\end{equation}}

\def\nn{\nonumber}

\begin{document}
\title{Space-time crystal and space-time group}
\author{Shenglong Xu}
\affiliation{Department of Physics, University of California,
San Diego, California 92093, USA}
\affiliation{Condensed Matter Theory Center and Department of Physics,
University of Maryland, College Park, MD 20742, USA}
\author{Congjun Wu}
\affiliation{Department of Physics, University of California,
San Diego, California 92093, USA}

\begin{abstract}
Crystal structures and the Bloch theorem play a fundamental role in
condensed matter physics.
We extend the static crystal to the dynamic ``space-time"
crystal characterized by the general intertwined space-time
periodicities in $D+1$ dimensions, which include both the static
crystal and the Floquet crystal as special cases.
A new group structure dubbed  ``space-time" group is constructed to
describe the discrete symmetries of space-time crystal.
Compared to space and magnetic groups, space-time group is augmented
by ``time-screw'' rotations and ``time-glide'' reflections involving
fractional translations along the time direction.
A complete classification of the 13 space-time groups in 1+1D is
performed.
The Kramers-type degeneracy can arise from the glide time-reversal
symmetry without the half-integer spinor structure, which constrains
the winding number patterns of spectral dispersions.
In 2+1D, non-symmorphic space-time symmetries enforce spectral
degeneracies, leading to protected Floquet semi-metal states.
Our work provides a general framework for further studying
topological properties of the $D+1$ dimensional space-time crystal.
\end{abstract}
\maketitle

The fundamental concept of crystal and the associated band theory based
on the Bloch theorem lay the foundation of condensed matter physics.
Studies on the crystal symmetry and band structure topology lead to the
discoveries of
topological insulators, topological superconductors, the Dirac
and Weyl semi-metal states \cite{Hasan2010,Qi2011,Chiu2016}.
Periodically driving further provides a new route to
engineer topological states even in systems originally
topologically trivial in the absence of driving, as
explored in the irradiated graphene \cite{Oka2009,Gu2011},
semiconducting quantum wells \cite{Lindner2011}, dynamically
modulated cold atom optical lattices \cite{Jotzu2014}, and
photonic systems \cite{Rechtsman2013a,Leykam2016}.
The periodicity of the quasi-energy enriches the topological
band structures \cite{Kitagawa2010,Asboth2014,Roy2016}, such as
the dynamically generated Majorana modes \cite{Thakurathi2013a},
1D helical channels \cite{Budich2017} and anomalous edge states
associated with zero Chern number  \cite{Rudner2013, Titum2016a}.
Topological classifications for interacting Floquet systems have
also been investigated \cite{Potter2016,Potter2016a,VonKeyserlingk2016a,
VonKeyserlingk2016b,Else2016}.

For periodically driven crystals, most studies treat the
temporal periodicity separately from the spatial one.
In fact, the driven system can exhibit much richer symmetry structures
than a simple direct product of spatial and temporal symmetries.
In particular, a temporal translation at a \textit{fractional}
period can be combined with the space group symmetries to form
novel space-time intertwined symmetries, which, to the best of
our knowledge, have not yet been fully explored.
For static crystals, the intrinsic connections between the space-group
symmetries and physical properties, especially the topological phases,
have been extensively studied \cite{Fu2011,Parameswaran2013, Young2015,
Wang2016a, Kruthoff2016, Watanabe2016, Bradlyn2017, Bouhon2017}.
Therefore, it is expected that the intertwined space-time symmetries
could also protect non-trivial properties of the driven system,
regardless of microscopic details.

In this article, we propose the concept of ``space-time'' crystal
exhibiting the intertwined space-time symmetries, whose periodicities
are characterized by a set of $D+1$ independent basis vectors, generally
space-time mixed.
The situation of separate spatial and temporal perodicities is
a special case and is also included.
The full discrete space-time symmetries of space-time crystals form a class
of new group structures -- dubbed the ``space-time'' group, which is
the generalization of space group by including ``time-screw'' and
``time-glide'' operations.
A complete classification of the 13 space-time groups in 1+1 D is
performed, and their constraints on band structure winding numbers
are studied.
In 2+1 D, 275 space-time groups are classified.
The non-symmorphic space-time symmetry operations, similar to
their static space-group counterparts, lead to the protected spectral
degeneracies for driven systems, even when the instantaneous
spectra are gapped at any given time.

\textit{Space-time crystal} --
We consider the time-dependent Hamiltonian $H=P^2/(2m)+V(\mathbf{r},t)$
in the $D+1$ dimensional space-time.
$V(\mathbf{r},t)$ exhibits the intertwined discrete space-time
translational symmetry as
\bea
V(\mathbf{r}, t)=V(\mathbf{r}+\mathbf{u}^i, t+\tau^i),
\ \ \ i=1,2, ..., D+1,
\label{eq:Ht}
\eea
where $(\mathbf {u}^i, \tau^i)=a^i$ is a set of the primitive basis
vectors.
In general, the space-time primitive unit cell is not a direct
product between spatial and temporal domains.
There may not even exist spatial translational symmetry at
any given time $t$, nor temporal translational
symmetry at any spatial location $\mathbf{r}$.
Consequently, the frequently used time-evolution operator
of one period for the Floquet problem generally does not apply.
The reciprocal lattice is spanned by the momentum-energy basis vectors
$b^i=(\mathbf{G^i}, \Omega^i)$ defined through
$b^i\cdot a^j= \sum_{m=1}^D G^i_m u^j_m -\Omega^i \tau^j =2\pi \delta^{ij}$.
The $D+1$ dimensional momentum-energy Brillouin zone (MEBZ)
may also be momentum-energy mixed.

\textit{Generalized Floquet-Bloch theorem}
We generalize the Floquet and Bloch theorems for the time-dependent
Schr\"odinger equation $i\hbar\partial_t \psi(\mathbf{r}, t)=
H(\mathbf{r},t)\psi(\mathbf{r}, t)$.
Due to the space-time translation symmetry, the lattice
momentum-energy vector $\kappa=(\mathbf{k},\omega)$ remains
conserved.
Only the $\kappa$ vectors inside the first MEBZ are non-equivalent,
and those outside are equivalent up to integer reciprocal lattice vectors.
The Floquet-Bloch states labeled by $\kappa$ take the form of
\bea
\psi_{\kappa,m}(\mathbf{r}, t)&=&e^{i (\mathbf{k}\cdot \mathbf{r}-
\omega_m t)}
u_m(\mathbf{r}, t),
\label{eq:gen_sol}
\eea
where $m$ marks different states sharing the common $\kappa$.
$u_m(\mathbf{r},t)$ processes the same space-time periodicity
as $H(\mathbf{r}, t)$, and is expanded as $u_m=\sum_{B} c_{m,B}
e^{i(\mathbf{G}\cdot \mathbf{r}-\Omega t)}$ with $B=(\mathbf{G}, \Omega)$
taking all the momentum-energy reciprocal lattice vectors.
The eigen-frequency $\omega_m$ is determined through the eigenvalue
problem defined as
\bea
\sum_{B^\prime}
\{ [\varepsilon_0(\mathbf{k+\mathbf{G}})-\Omega]
\delta_{B,B'}+V_{B-B'} \} c_{m,B^\prime} 
=\omega_m c_{m,B}, \ \ \, \ \ \,
\label{eq:eigen_nonint}
\eea
where $\varepsilon_0(\mathbf{k})$ is the free dispersion, and
$V_{B}$ is the momentum-energy Fourier component of the
space-time lattice potential $V(\mathbf{r},t)$.
The dispersion based on Eq. \ref{eq:eigen_nonint} is represented
by a $D$-dimensional surface in the MEBZ which is a $D$+1
dimensional torus.

\begin{figure}
\includegraphics[height=0.3\columnwidth,width=0.49\columnwidth]
{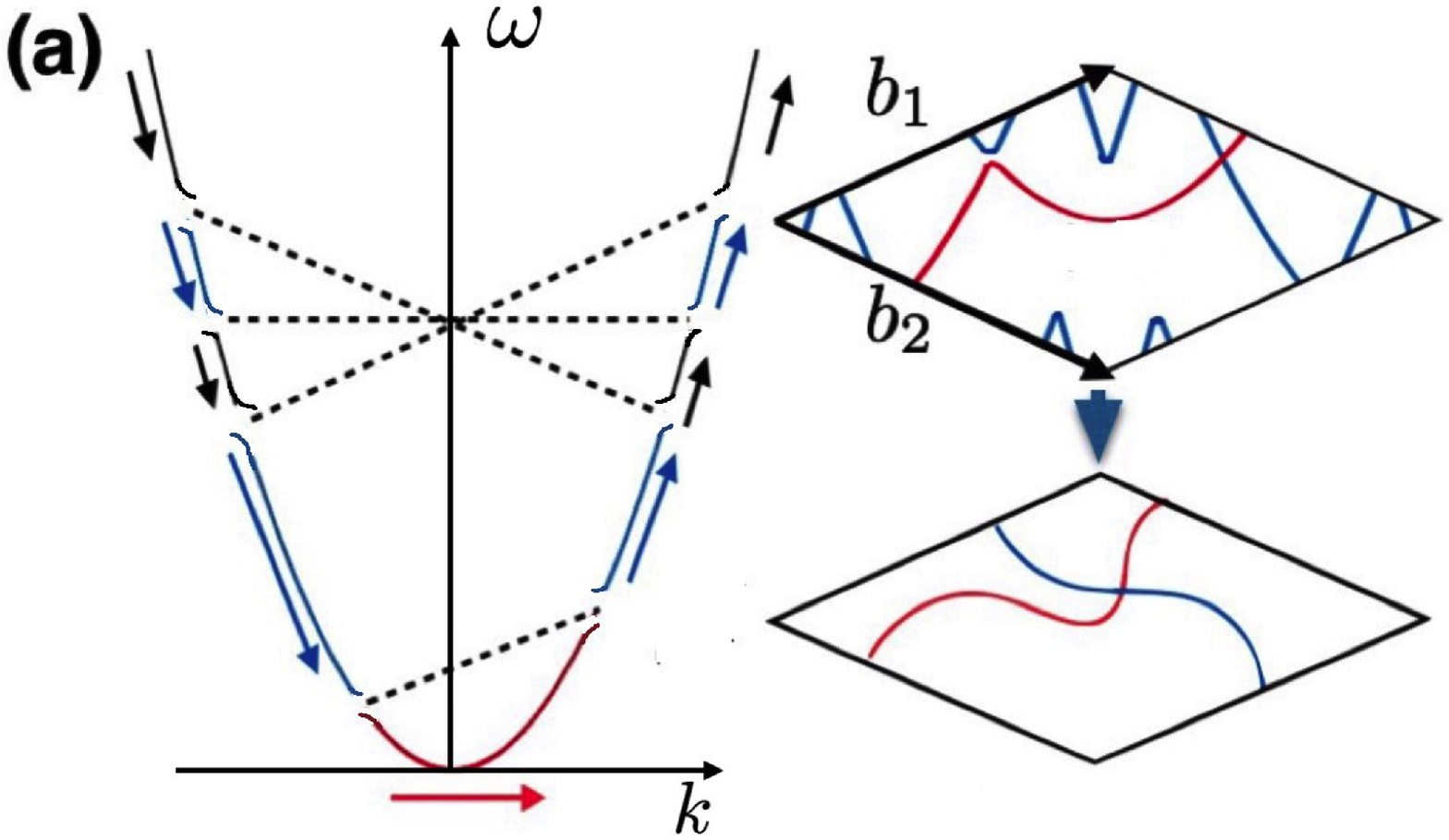}
\includegraphics[height=0.3\columnwidth,width=0.49\columnwidth]
{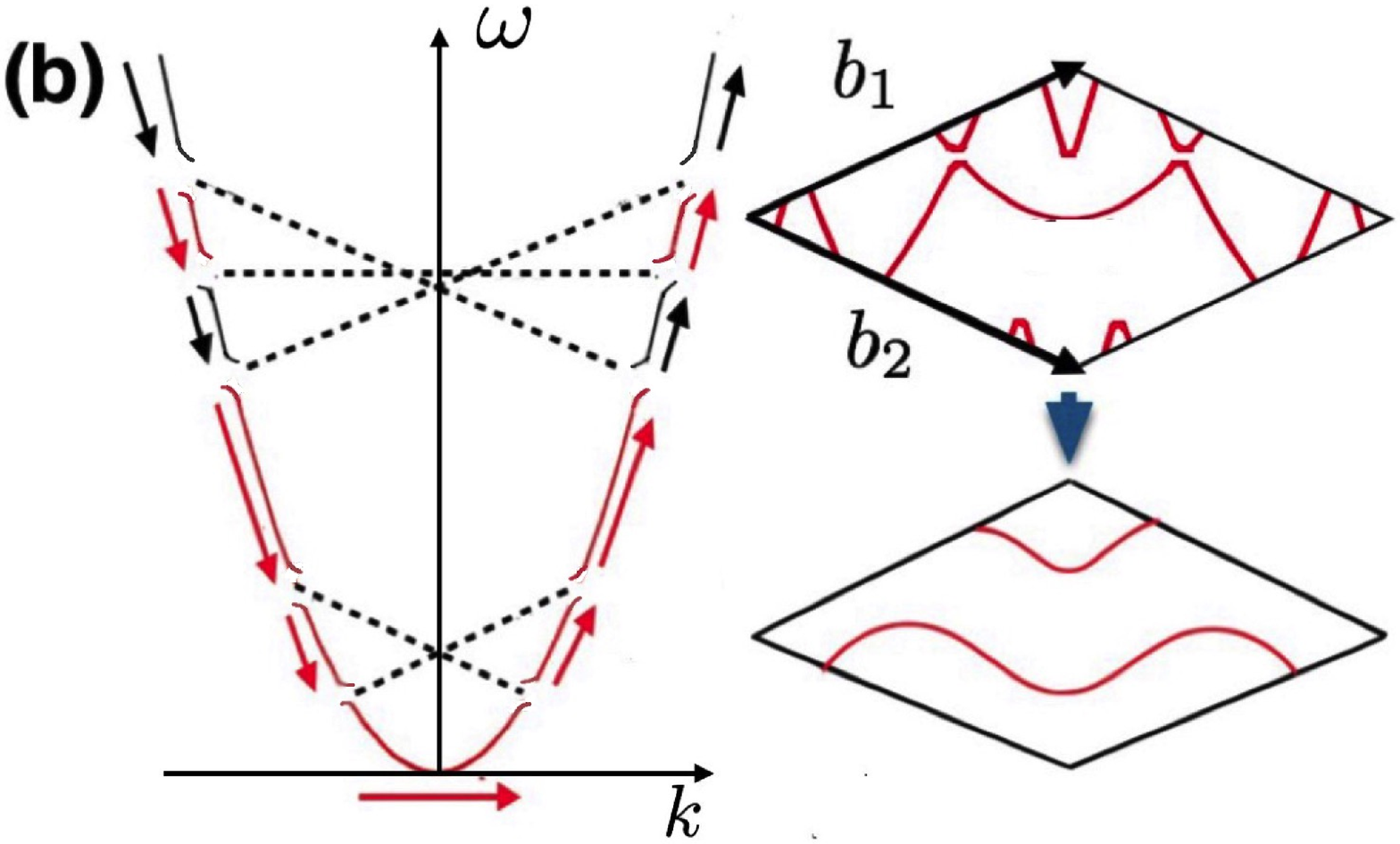}
\caption{Folding the band dispersions of the 1+1 D space-time crystal
into the 1st rhombic MEBZ in the weak lattice limit.
The momentum-energy reciprocal lattice vectors of nonzero $V_B$'s are
represented by dashed lines.
The low-energy part of the free dispersion curve evolves to closed
loops.
($a$) Two loops with the winding numbers $\mathbf{w_r}=(1,0)$ (red)
and $\mathbf{w_b}=(0,1)$ (blue).
($b$) An extra nonzero $V_G$ connects two loops in ($a$) forming a new
one with $\mathbf{w}=\mathbf{w_r}+\mathbf{w_b}$.
}
\label{fig:insulator}
\end{figure}

\textit{Dispersion winding numbers} --
The band structure of the space-time crystal exhibits novel
features different from those of the static crystal.
For simplicity, below we use the 1+1 D case for an illustration.
The dispersion relation $\omega(k)$ forms closed loops in the 2D
toroidal MEBZ, each of which is characterized by a pair of
winding numbers $\mathbf{w}=(w_1,w_2)$.
Compared to the static case in which the band dispersion only
winds around the momentum direction, here $\omega(k)$ is typically
not single-valued and its winding patterns are much richer.
The dispersions in the limit of a weak space-time potential
$V(x,t)$ with a rhombic MEBZ are illustrated in
Fig. \ref{fig:insulator} ($a$) and ($b$), with details
presented in Supplemental Material (S.M.) Sect. A \cite{supp}.
When folded into the MEBZ, the free dispersion curve $\varepsilon_0(k)$
can cross at general points not just on high symmetry ones.
A crossing point corresponds to two equivalent momentum-energy
points related by a reciprocal vector $G$.
When $V_G\neq 0$, the crossing is avoided by forming a gap at
the magnitude of $2|V_G|$.
The total number of states at each $\mathbf{k}$ is independent of
the strength of $V(x,t)$, hence crossing can only split along the
$\omega$-direction and $d\omega/dk$ is always finite.
Consequently, trivial loops with the winding numbers $(0,0)$
are forbidden.
Generally, the winding directions of the dispersion loops are
momentum-energy mixed.
Furthermore, different momentum-energy reciprocal lattice vectors can
cross each other, leading to composite loops  winding around
the MEBZ along both directions as shown in Fig. \ref{fig:insulator} ($b$).
Hence, in general all patterns $(w_1,w_2)$ are possible
except the contractible loops.

{\it Space-time group --}
To describe the symmetry properties of the $D+1$ dimensional
space-time crystals, we propose a new group structure dubbed
``space-time'' group defined as the discrete subgroup of the
direct product of the Euclidean group in $D$ spatial dimensions
and that along the time-direction $E_D \otimes E_1$.
Please note that in general the space-time group cannot be factorized
as the direct product between discrete spatial and temporal subgroups.
It not only includes space and magnetic group transformations
in the $D$-spatial dimensions, but also includes
operations involving fractional translations along the time-direction.
Since space and time are non-equivalent in the Schr\"odinger equation,
space-time rotations are not allowed except the 2-fold case.

To be concrete, a general space-time group operation $\Gamma$
on the space-time vector $(\mathbf{r},t)$ is defined as,
\bea
\Gamma(\mathbf{r},t)=(R\mathbf{r} +\mathbf{u}, st+\tau),
\eea
where $R$ is a $D$-dimensional point group operation, $s=\pm 1$
and $s=-1$ indicates time-reversal,
and $(\mathbf{u},\tau) =\sum_i m_i a^i$ represents a space-time
translation with $m_i$ either integers or fractions.
If $\tau=0$, $\Gamma$ is reduced to a space group or magnetic group
operation according to $s=\pm 1$, respectively.
If $\tau\neq 0$, when $(\mathbf{u}, \tau$) contains fractions of $a^i$,
new symmetry operations arise due to the dynamic nature of
the crystal potential,
including the ``time-screw'' rotation and ``time-glide'' reflection,
which are a spatial rotation and a reflection followed by a fractional time
translation, respectively.
The operation of $\Gamma$ on the Hamiltonian
is defined as $\Gamma^{-1} H(\mathbf{r},t) \Gamma = H(\Gamma(\mathbf{r},t))$,
or, $\Gamma^{-1} H(\mathbf{r},t) \Gamma= H^*(\Gamma (\mathbf{r},t))$
for $s=\pm 1$, respectively.
Correspondingly, the transformation $M_\Gamma$ on the Bloch-Floquet
wavefunctions $\psi_\kappa(\mathbf{r},t)$ is
$M_\Gamma \psi_\kappa=\psi_\kappa(\Gamma^{-1}(\mathbf{r},t))$, or,
$\psi_\kappa^*(\Gamma^{-1}(\mathbf{r},t))$ for $s=\pm 1$, respectively.

\begin{figure}
\includegraphics[height=0.85\columnwidth,width=0.8\columnwidth]{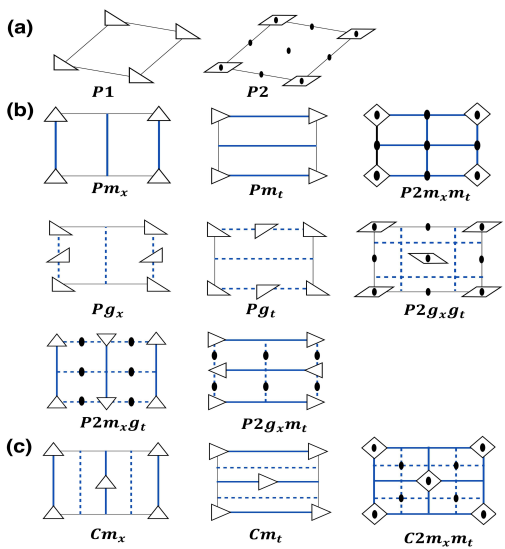}
\caption{The classification of 13 space-time groups in 1+1D and
the associated crystal configurations.
The solid oval marks the 2-fold space-time axis, and the parallelogram
means the 2-fold axis without reflection symmetries.
The thick solid and dashed lines represent reflection and
glide-reflection axes, respectively.
Configurations of triangles and the diamond denote the local
symmetries under reflections.
($a$) The oblique lattices with and without 2-fold axes.
Their basis vectors are generally space-time mixed.
The primitive ($b$) and centered ($c$) orthorhombic lattices:
According to their reflection and glide reflection symmetries,
they are classified to 8 groups in ($b$), and
3 groups in ($c$).}
\label{fig:sptm_group}
\end{figure}

Now we present a complete classification of the 1+1 D space-time groups.
Due to the non-equivalence between spatial and temporal directions,
there are no square and hexagonal space-time crystal systems.
The point-group like operations are isomorphic to $D_2$, including
reflection $m_x$, time reversal $m_t$, and their combination
$m_xm_t$, {\it i.e.}, the 2-fold space-time rotation.
Consequently, only two space-time crystal systems are allowed --
oblique and orthorhombic.
There exist two types of glide reflections: the time-glide
reflection $g_x$,
and $g_t$ denoted as ``glide-time-reversal'' is
time-reversal followed by a fractional
translation along the $x$-direction.

The above 1+1 D space-time symmetries give rise to 13 space-time groups
in contrast to the 17 wallpaper space groups characterizing the
2D static crystals.
The oblique Bravais lattice is simply monoclinic, while the orthorhombic
ones include both the primitive and centered Bravais lattices.
The monoclinic lattice gives rise to two different crystal
structures with and without the 2-fold space-time axes, whose
space-time groups are denoted by $P_{1,2}$, respectively,
as shown in Fig. \ref{fig:sptm_group} (a).
For the primitive orthorhombic lattices, the associated crystal
structures can exhibit the point-group symmetries $m_x$ and $m_t$, and
the space-time symmetries $g_t$ and $g_x$.
Their combinations give rise to 8 space-time crystal structures
denoted as $Pm_x$, $Pm_t$, $P2m_xm_t$, $Pg_x$, $Pg_t$, $P2g_xg_t$,
$P2m_xg_t$, $P2g_xm_t$, respectively, as shown in Fig. \ref{fig:sptm_group} (b).
Four of them possess the 2-fold space-time axes as indicated
by ``2'' in their symbols.
For the centered orthorhombic Bravais lattices, 3 crystal structures
exist with space-time groups denoted as $Cm_x$, $Cm_y$, and $C2m_xm_t$,
respectively, as shown in Fig. \ref{fig:sptm_group} (c).
They all exhibit glide-reflection symmetries, and the last one
possesses the 2-fold space-time axes as well.
Two unit cells are plotted for the centered lattices to show
the full symmetries explicitly, and their primitive basis
vectors are actually space-time mixed.

The classifications of the space-time groups in higher dimensions are
generally complicated.
A general method is the group cohomology as presented in
Sect B of S. M. \cite{supp}.
In particular, the classification of 2+1D space-time group is outlined
in Sect C of S. M. \cite{supp}, whose structures are further enriched
by spatial rotations and time-screw rotations.
Compared to the 3D static crystals, the cubic crystal systems are
not allowed, and two different monoclinic crystal systems appear
with the perpendicular axis along the time and spatial directions,
respectively.
In total, there are 7 crystal systems and 14 Bravais lattices,
but 275 space-time groups.

\begin{figure}
\includegraphics[height=0.45\columnwidth, width=0.4\columnwidth]
{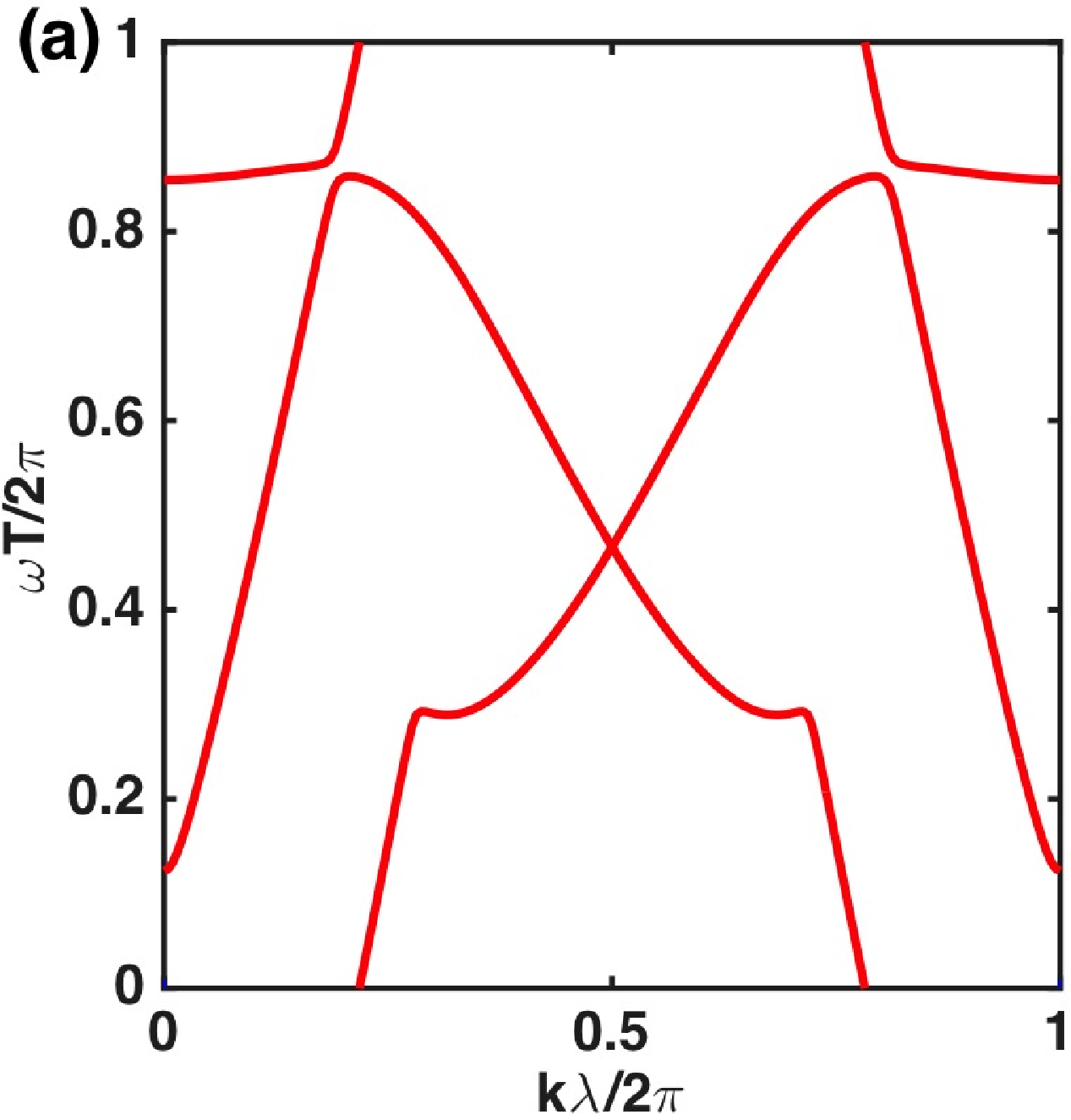}
\includegraphics[height=0.45\columnwidth, width=0.4\columnwidth]
{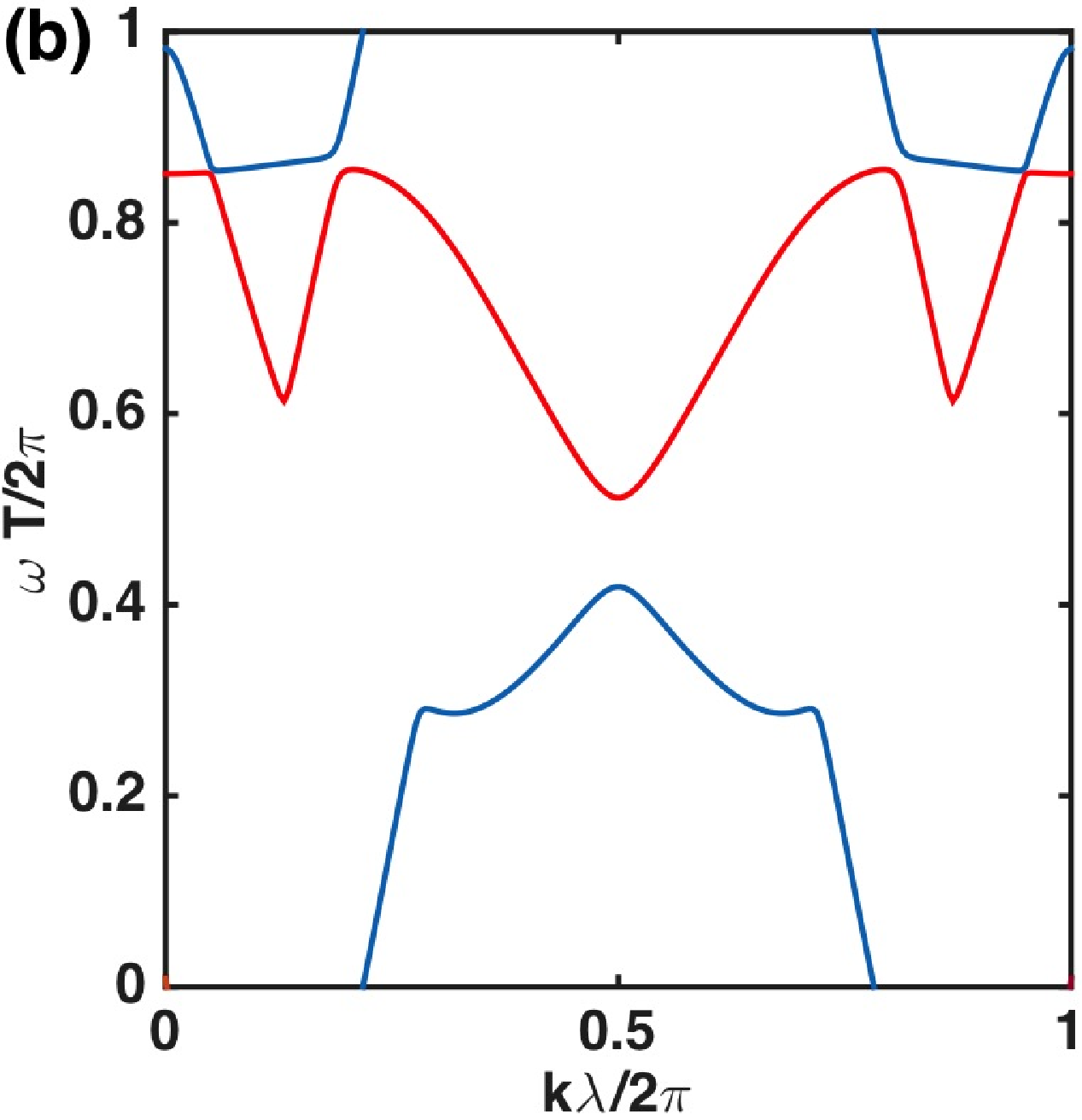}
\caption{(a) The Floquet-Bloch band spectrum with the space-time
  lattice potential possessing the glide time-reversal
  symmetry $g_t$.
  When applied to the states with $\kappa_x=\pi/\lambda$, $g_t$
  becomes a Kramers symmetry protecting the double-degeneracy.
  (b) Lifting the Kramers degeneracy by adding
a glide time-reversal symmetry breaking term.}
\label{fig:gt}.
\end{figure}

\textit{Protection of spectral degeneracy}
The intertwined space-time symmetries besides translations can protect
spectral degeneracies.
Below we consider the effects from the Kramers symmetry without spin
and the non-symmorphic symmetries for the 1+1 D and 2+1 D space-time
crystals, respectively.

Consider a 1+1 D space-time crystal whose unit cell is a direct product
of spatial and temporal periods $\lambda$ and $T$, respectively.
We assume the system is invariant under the glide time-reversal operation
$g_t(x,t) =(x+\frac{1}{2}\lambda, -t)$, whose operation on the Hamiltonian
is defined as $g_t^{-1} H g_t = H^*(g_t(x,t))$.
The corresponding transformation $M_{g_t}$ on the Bloch-Floquet
wavefunction $\psi_{\kappa}(x,t)$ of Eq. \ref{eq:gen_sol} is
anti-unitary defined as $M_{g_t}\psi_{\kappa}=
\psi^*_\kappa(g_t^{-1}(x,t))$.
This glide time-reversal operation leaves the line of
$\kappa_x=\pi/\lambda$ in the MEBZ invariant.
$M_{g_t}$ becomes a Kramers symmetry for states with $\kappa_x=\pi/\lambda$,
\bea
M_{g_t}^2 \psi_{\kappa}= \psi_{\kappa} (x-\lambda,t)=
e^{-i\kappa_x\lambda}\psi_{\kappa}=-\psi_{\kappa},
\eea
without involving the half-integer spinor structure.
It protects the double degeneracy of the momentum-energy
quantum numbers of $\psi_{\kappa}$ and $M_{g_t}\psi_{\kappa}$.
Hence the crossing at $\kappa_x=\pi/\lambda$ cannot be avoided
and the dispersion winding numbers along the momentum direction
must be even.

As a concrete example, we study a crystal potential with the above
spatial and temporal periodicities,
$V(x,t)=V_0 \big( \sin \frac{2\pi}{T} t \cos \frac{2\pi}{\lambda}x+
\cos \frac{2\pi}{T}t \big)$.
Except the glide time-reversal symmetry, it does not possess other symmetries.
Its Bloch-Floquet spectrum is calculated based on Eq. \ref{eq:eigen_nonint},
and a representative dispersion loop is plotted in the MEBZ
shown in Fig. \ref{fig:gt} (a).
The crossing at $\kappa_x=\pi/\lambda$ is protected by the glide
time-reversal symmetry giving rise to a pair of Kramers doublet.
As a result, the winding number of this loop is
$\mathbf{w}=(w_x,w_t)=(2,0)$.
If a glide time-reversal breaking term
$\delta V= V_0^\prime \cos(\frac{2\pi}{\lambda}x)$ is added
into the crystal potential, the crossing is avoided
as shown in Fig. \ref{fig:gt} (b).
Consequently, the dispersion splits into two loops,
both of which exhibit the winding number $(1,0)$.
Similarly, out of the 8 primitive orthorhombic space-time crystals,
3 of them, $Pg_t$, $P2g_xg_t$, and $P2g_tm_x$, enforce this non-spinor
type Kramers degeneracy, while the other 5
generally does not protect such a degeneracy.

\begin{figure}
\includegraphics[height=0.35\columnwidth, width=0.35\columnwidth]
{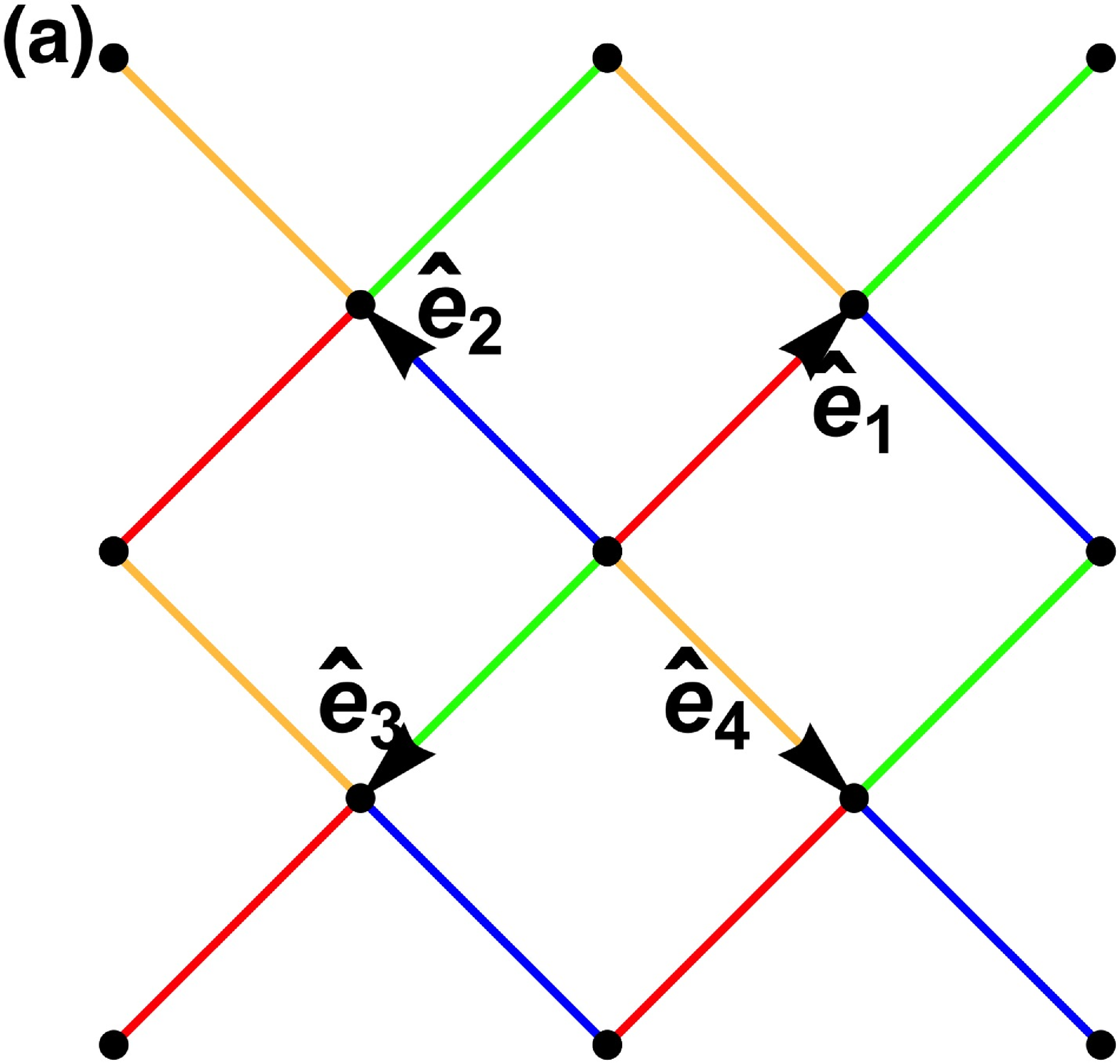}
\includegraphics[height=0.35\columnwidth, width=0.35\columnwidth]
{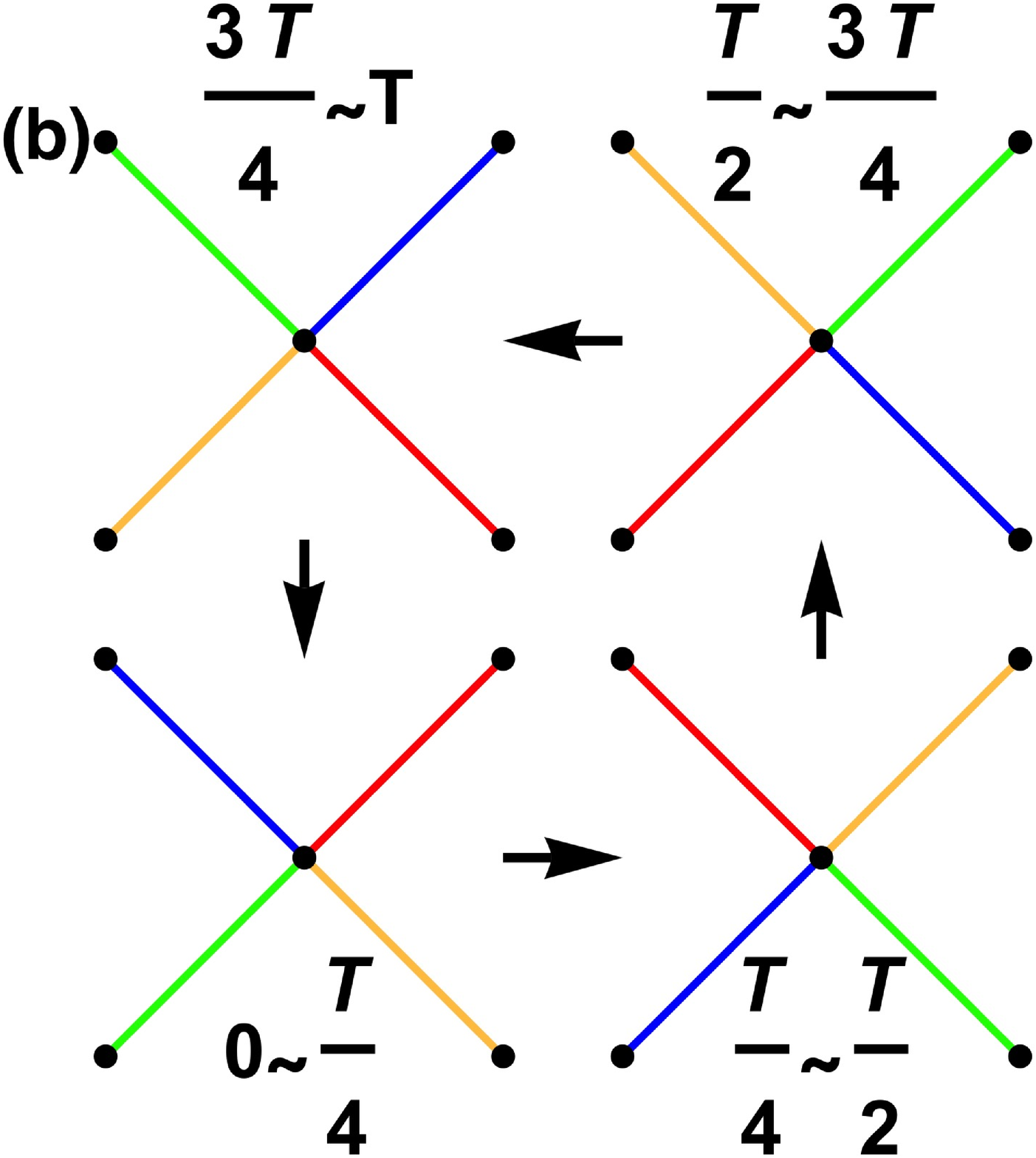}
\includegraphics[height=0.35\columnwidth, width=0.35\columnwidth]
{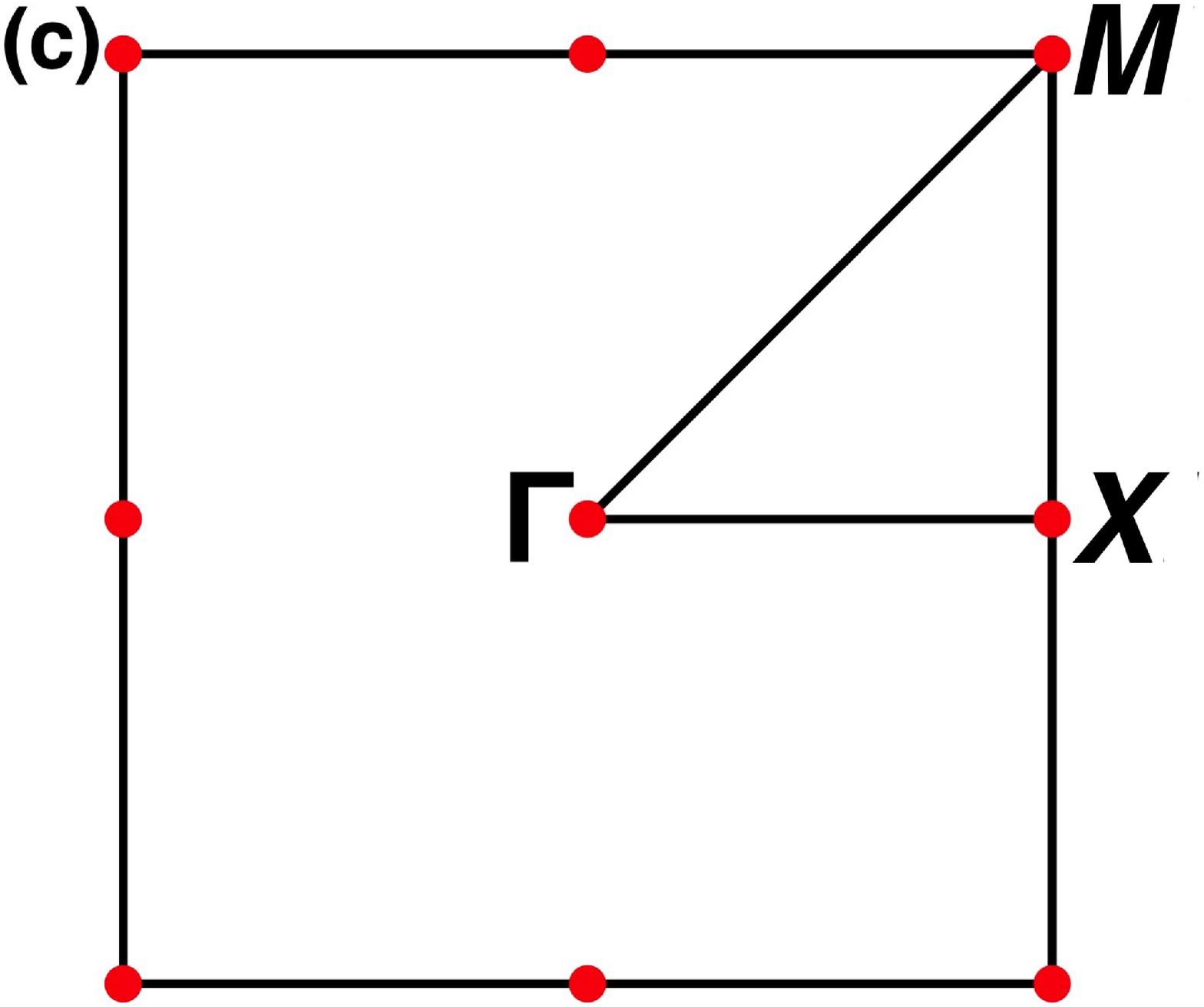}
\includegraphics[height=0.35\columnwidth, width=0.35\columnwidth]
{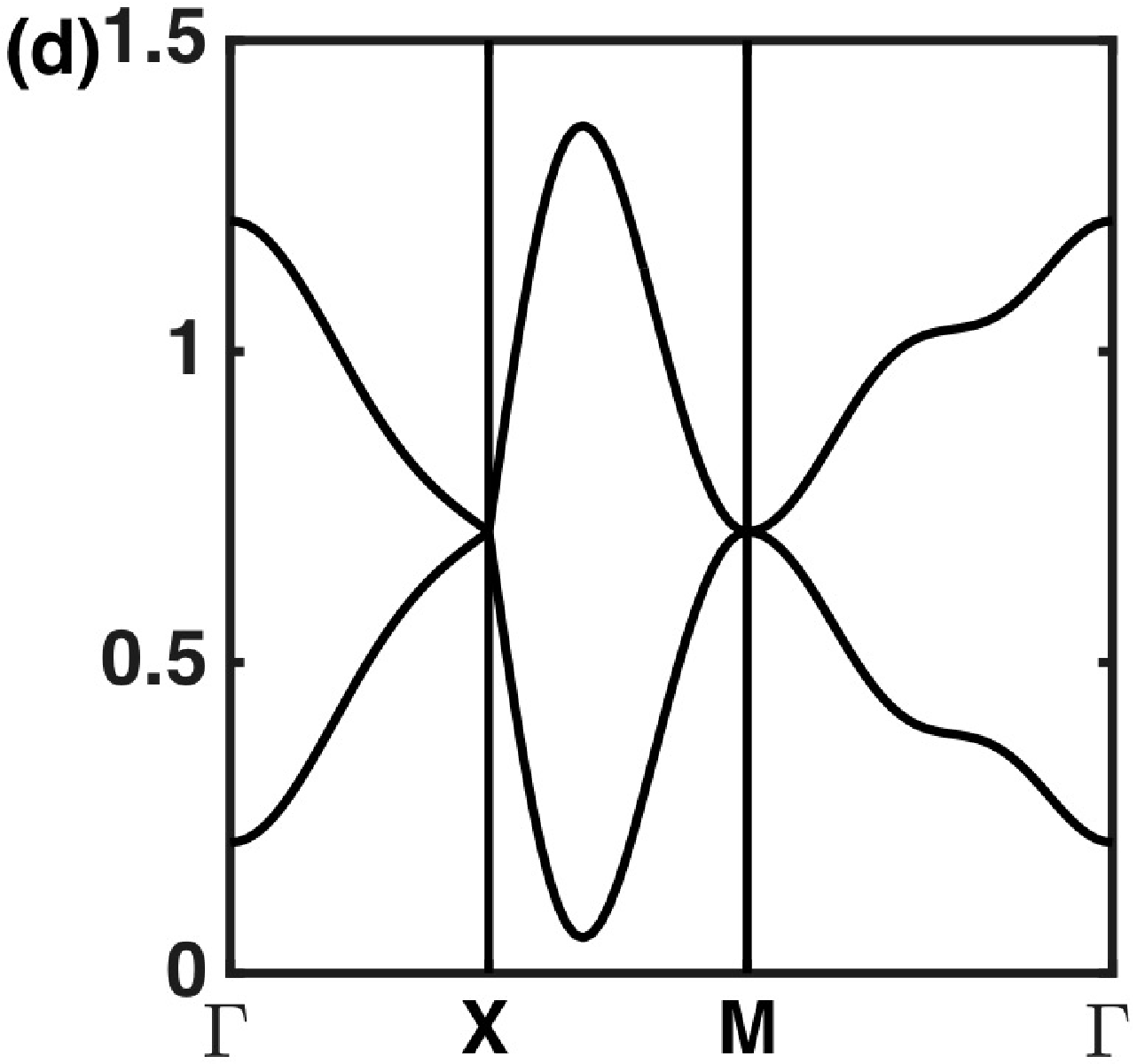}
\caption{\small ($a$) The 2+1 D space-time lattice structure of the
Hamiltonian Eq. \ref{eq:floquet}.
The bond directions are marked as $\vec e_{1,3}=\pm \frac{1}{2}
(\hat x +  \hat y)$, $\vec e_{2,4}=\mp \frac{1}{2}(\hat x -  \hat y)$.
($b$) The time-dependent hopping pattern rotates 90$^\circ$ every one
quarter period.
The bonding strengths $w_{e_i}(t)$ of the $R$, $B$, $G$ and $Y$ bonds
equal $0.2$, $3$, $-3.2$, and $0.5$, respectively.
($c$) The momentum Brillouin zone with high symmetry points $\Gamma=(0,0)$,
$M=(\pm\pi,\pm\pi)$, and $X=(0,\pm \pi)$ and $(\pm\pi,0)$.
($d$) The dispersions along the cuts from $\Gamma$ to
$X$ to $M$ to $\Gamma$.
Two-fold degeneracies appear at $X$ and $M$.
}
\label{fig:2Ddrv}
\end{figure}

Next we present a 2+1 D Floquet semi-metal state, whose spectral
degeneracies are protected by non-symmorphic space-time group
operations.
Consider that the space-time little group of the momentum $\mathbf{k}$
contains two non-symmorphic space-time group operations $g_{1,2}$,
both of which do not flip the time direction, hence, they are represented
by unitary operators.
If they satisfy
\bea
g_1 g_2 =T g_2 g_1,
\eea
where $T$ is a translation of integer lattice vectors.
As shown in Sect. D in S. M. \cite{supp}, $T$ can only be
a spatial translation without involving the time denoted as
$T(\mathbf{u})$.
Assume $\mathbf{k} \cdot \mathbf{u}=2\pi p/q$ with $p$ and $q$ coprime,
we find that the Bloch-Floquet wavefunctions exhibit a $q$-fold
degeneracy at the momentum-energy vector $\kappa=(\mathbf{k},\omega)$
proved  as follows.
Since $g_1$ belongs to the little group,
$\psi_{\kappa}(\mathbf{r},t)$ can be
chosen to satisfy $M_{g_1} \psi_{\kappa,1}=\mu \psi_{\kappa,1}$,
then $\psi_{\kappa}, M_{g_2}\psi_{\kappa},
M_{g_2}^2\psi_{\kappa}, ...., M_{g_2}^{q-1}\psi_{\kappa}$ are the common
Bloch-Floquet eigenstates sharing the same $\kappa$ but exhibiting
a set of different eigenvalues of $g_1$ as $\eta, \mu \eta, \mu\eta^2,
..., \mu \eta^{q-1}$ with $\eta=e^{i\pi p/q}$.
Then they are orthogonal to each other forming a $q$-fold degeneracy.
Compared to the case of non-symmorphic space group protected
degeneracy \cite{Parameswaran2013, Young2015, Watanabe2016},
here $g_{1,2}$ are space-time operations for a
dynamic space-time crystal.
For the case that one or both of $g_{1,2}$ flip the time direction,
the situation is more involved due to involving anti-unitary
operations.
Protected degeneracies are still possible as presented in
Sect. D in S. M. \cite{supp}.

We employ a 2+1 D tight-binding space-time model as an example
to illustrate the above protected degeneracy.
A snap shot of the lattice is depicted in Fig. \ref{fig:2Ddrv} ($a$),
which consists of two sublattices:
The $A$-type sites are with integer coordinates $(i,j)$, and each
$A$-site emits four bonds along $\vec e_{i}$ to its four
neighboring $B$ sites at $(i\pm\frac{1}{2},j\pm\frac{1}{2})$.
The space-time Hamiltonian within the period $T$ is
\bea
H (t) = - \sum_{\vec r\in A, ~~ \vec r+\frac{a}{2}\vec e_i \in B}
\big \{ w_{\vec e_i} (t) c^\dagger_{\vec r} d_{\vec r+ \frac{a}{2}\vec e_i}
+ h.c., \big\},
\label{eq:floquet}
\eea
where $a$ is the distance between two nearest $A$ sites, and
$w_{\vec e_i}(t)$'s are hopping amplitudes with different strengths.
Their time-dependence is illustrated in Fig. \ref{fig:2Ddrv} ($b$):
Within each quarter period, $w_{\vec e_i}$ does not vary,
and their pattern rotates 90$^\circ$ after every $T/4$.
At each given time, the lattice possesses a simple 2D space group symmetry
$p2111$, which only includes two-fold rotations
around the $AB$-bond centers without reflection and glide-plane symmetries.
For example, the rotation $R_\pi$ around $(\frac{a}{4},\frac{a}{4})$
transforms the coordinate $(x,y,t) \to (\frac{a}{2}-x,\frac{a}{2}-y, t)$.
In addition, there exist ``time-screw'' operations, say,
an operation $S$ defined as a rotation around an $A$-site
$(0,0)$ at 90$^\circ$ followed by a time-translation at $T/4$,
which transforms $(x,y,t) \to (y,-x, t+\frac{T}{4})$.
$R_\pi$ and $S$ are generators of the space-time group for
the Hamiltonian Eq. \ref{eq:floquet}.
Since $S$ is a time-screw rotation, this space-time group is
non-symmorphic.
It is isomorphic to the 3D space-group $I4_1$, while its 2D
space subgroup $p2111$ is symmorphic.
We have checked that, for a static Hamiltonian taking any of the
bond configuration in Fig. \ref{fig:2Ddrv} ($b$), the
energy spectra are fully gapped.
However, the non-symmorphic space-time group gives rise to
spectral degeneracies.
Its momentum Brillouin zone is depicted in Fig. \ref{fig:2Ddrv} ($c$).
The space-time little group of the $M$-point $(\pi,\pi)$ contains
both $R$ and $S$ satisfying $RS=T(a\hat y)SR=-SR$.
Similarly, the $X$-point $(\pi,0)$ is invariant under both $R$ and $S^2$
satisfying $RS^2=T(a\hat x +a \hat y)S^2R= -S^2R$.
Hence, the Floquet eigen-energies are doubly degenerate
at $M$ and $X$-points as shown in Fig. \ref{fig:2Ddrv} ($d$),
showing a semi-metal structure.

In conclusion, we have studied a novel class of $D+1$ dimensional
dynamic crystal structures exhibiting the general space-time periodicities.
Their MEBZs are $D+1$ dimensional torus and are typically momentum-energy
entangled.
The band dispersions exhibit non-trivial windings around the MEBZs.
The space-time crystal structures are classified by space-time
group, which extend space group for static crystals
by incorporating time-screw rotations and time-glide
reflections.
In 1+1D, a complete classification of the 13 space-time groups is
performed, and there exist 275 space-time groups in 2+1 D.
Space-time symmetries give rise to novel Kramers degeneracy
independent of the half-integer spinor structure.
The non-symmorphic space-time group operations lead to protected
spectral degeneracies for space-time crystals.
This work sets up a symmetry framework for exploring novel properties of space-time crystals.
It also serves as the starting point for future studies, for example, the
dynamical topological phases of matter based on their space-time groups.

{\it Acknowledgments}
This work is supported by AFOSR FA9550-14-1-0168.

{\it Note added}.
Upon completing this manuscript, we noticed an interesting and important
work by T. Morimoto et. al. \cite{Morimoto2017} classifying Floquet topological
crystalline insulators with two-fold space-time symmetries.

\appendix
\section{The nearly-free-particle approximation of 1+1 D space-time crystal}

\begin{figure}
\includegraphics[height=0.45\columnwidth, width=0.45\columnwidth] 
{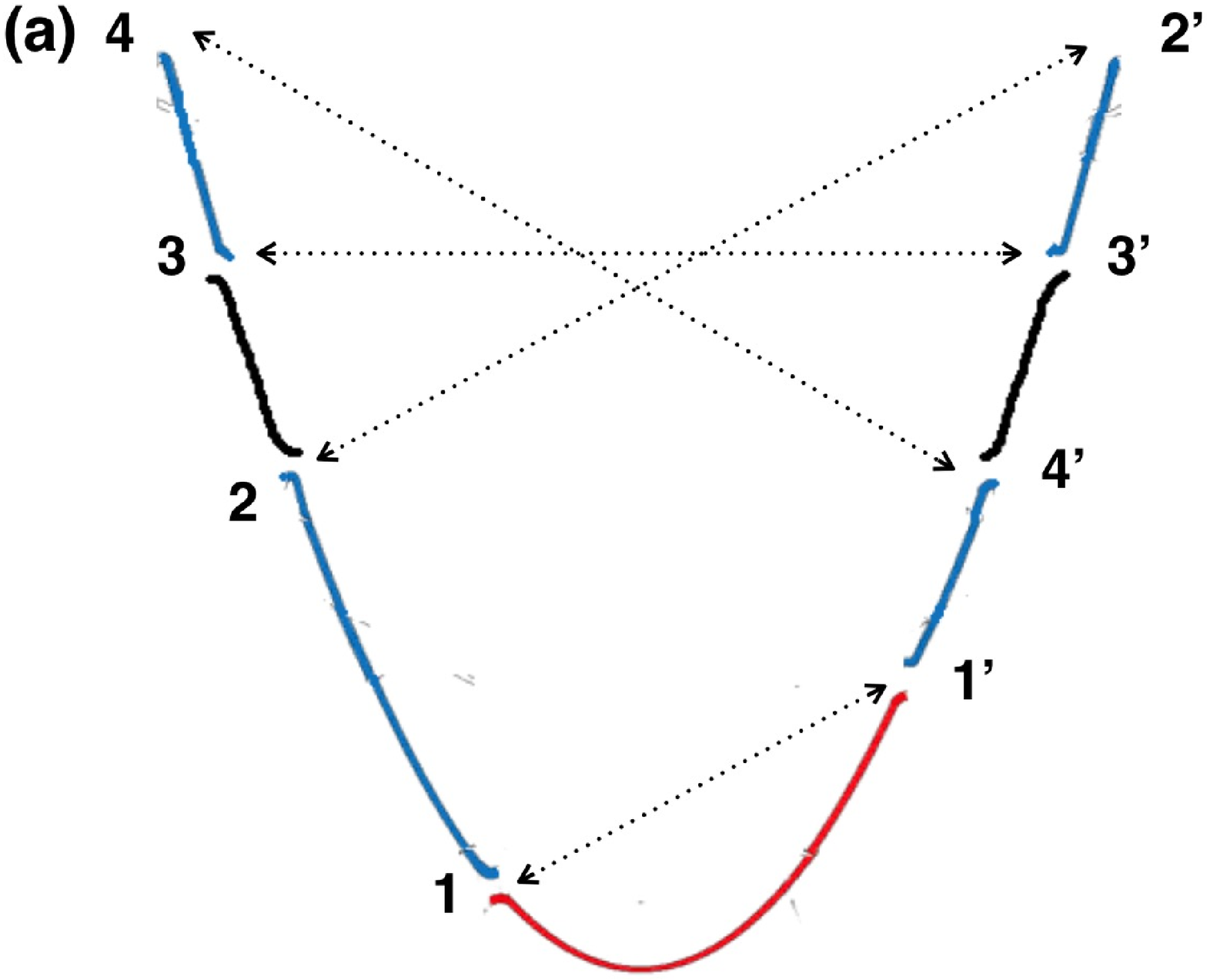}
\includegraphics[height=0.45\columnwidth, width=0.45\columnwidth] 
{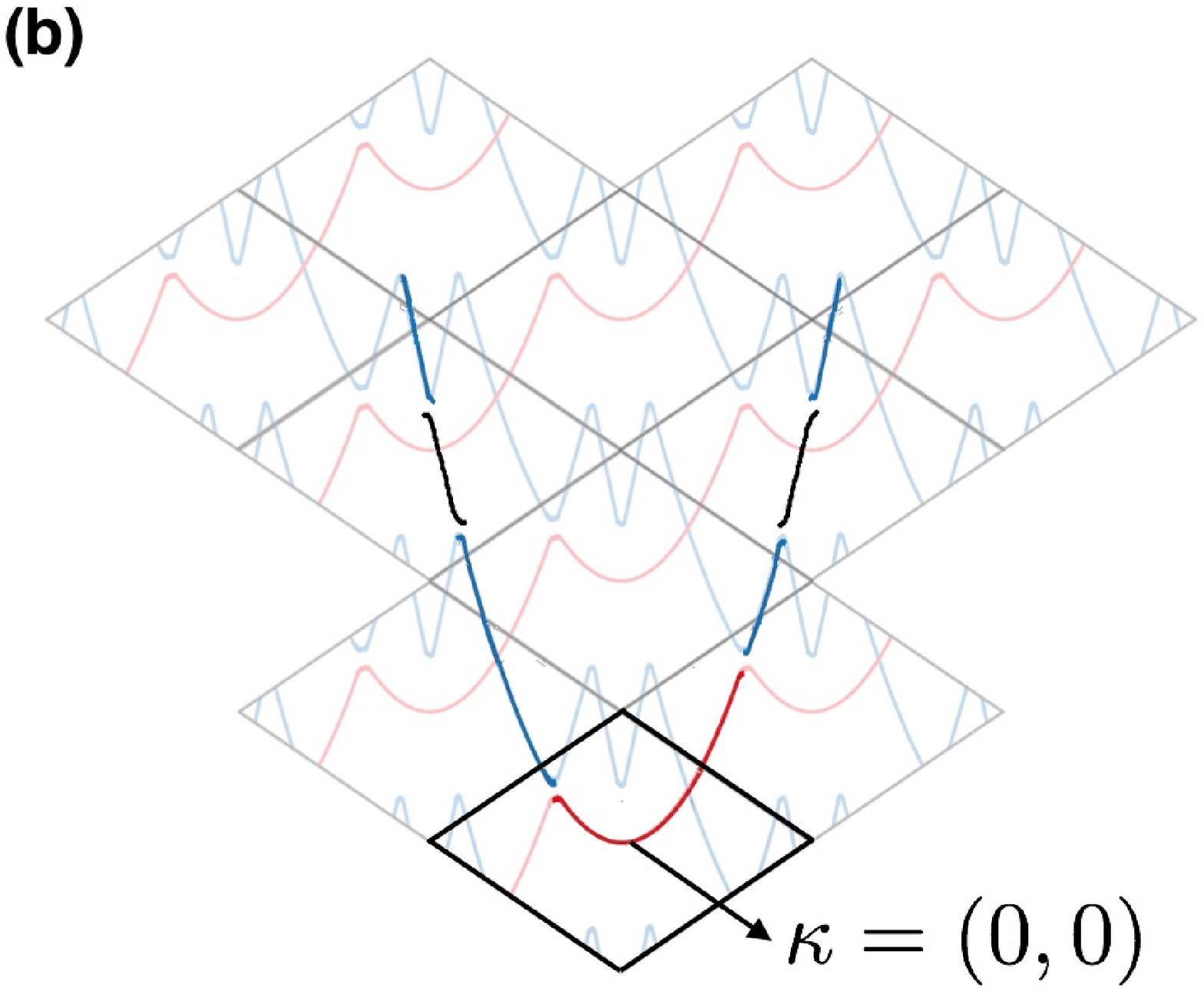}
\includegraphics[height=0.5\columnwidth, width=0.6\columnwidth] 
{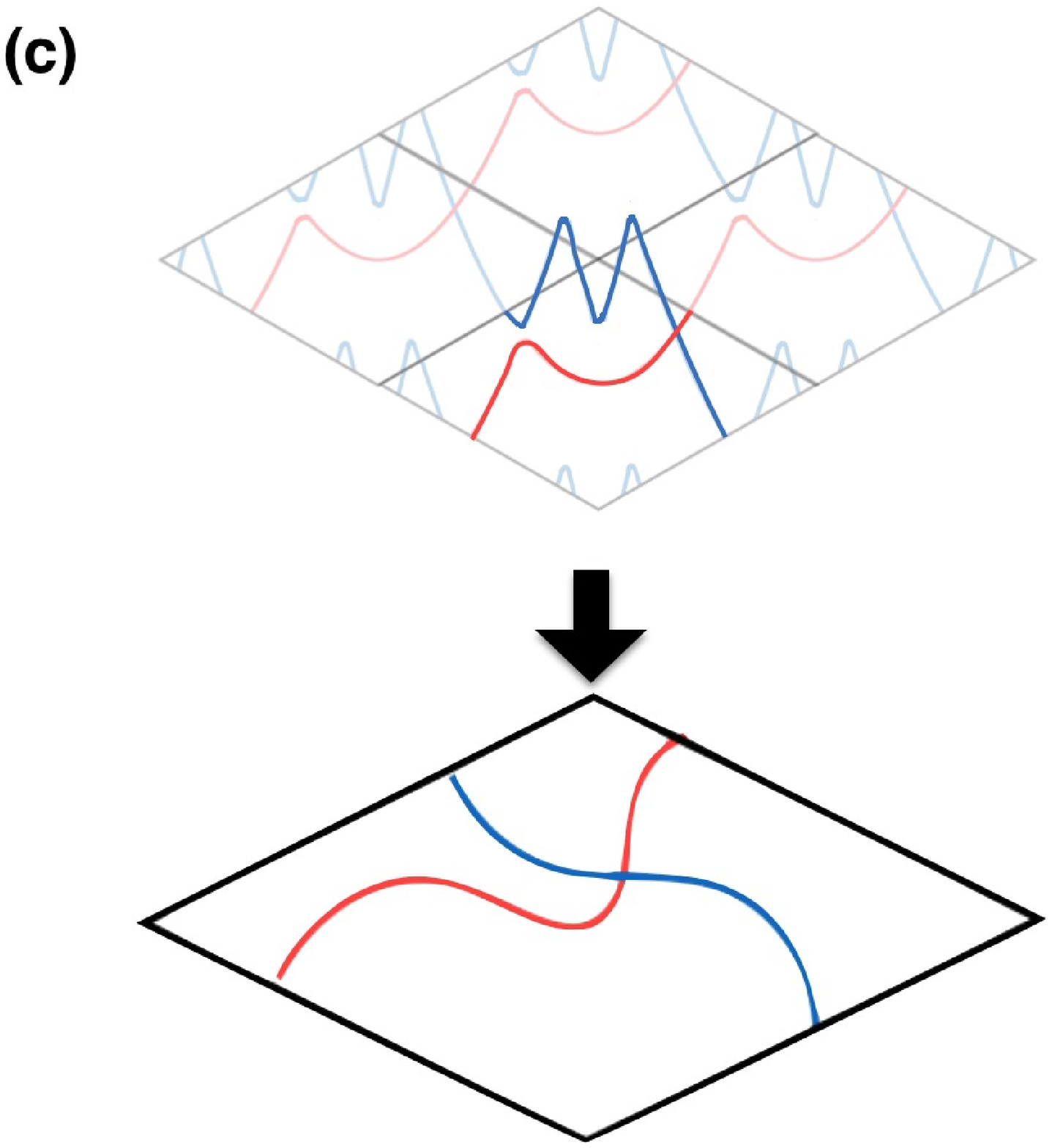}
\caption{(a). The nearly-free-particle dispersion without the
zone-folding.
When two momentum-energy vectors are differed by a reciprocal lattice
vector $B$, whose Fourier component of the space-time lattice
potential $V_B\neq 0$, they are marked in pairs.
$V_B$ results in level repulsion and open a gap at the value of $2|V_B|$.
(b) The dispersion duplicated in extended MEBZs.
The relevant MEBZs are sketched in the background for clarity.
The first MEBZ is centered on the origin.
(c) We smoothly vary the loop structure so that the winding
number structure becomes transparent.
}
\label{fig:zonefolding}
\end{figure}

In this section, we expand the discussion on  the nearly-free-particle
band structure of the space-time crystal and explicitly demonstrate the
momentum-energy Brillouin zone (MEBZ) folding procedure.
We consider a weak space-time lattice potential
$V(\mathbf{r}, t)$ and work in the framework of the generalized
Floquet-Bloch theorem based on Equations (2) and (3) in the main text.

Consider a Floquet-Bloch state $\psi_{\kappa,m}(\mathbf r, t)$ with
the good quantum number $\kappa=(\mathbf{k},\omega_m)$ and the band
index $m$.
Its wavefunction component in terms of each reciprocal lattice vector
$B=(\mathbf G, \Omega)$ is denoted as $c_{m,B}$.
By construction, if $c_{m, B}$ is a solution, then $c_{m,B-B\prime}$
with $B^\prime=(\mathbf{G^\prime}, \Omega^\prime)$ corresponds to the
solution with an equivalent quasi-momentum and energy
$(\mathbf{k}+\mathbf{G}', \omega_m+\Omega')$.
Nevertheless, this remains the same state as before. 
In order to remove this redundancy, $\kappa$ can be constrained 
in the first momentum-energy Brillouin zone (FMEBZ), i.e., the unit cell
in the reciprocal space centered at the origin.
Dispersions residing in high order MEBZs can be folded into
the FMEBZ. 
The general process is as follows:
First fold the free spectrum $\varepsilon_0(\mathbf k)$ into the FMEBZ.
When it crosses, the corresponding $\kappa$'s of the unfolded spectrum
are differed by a momentum-energy reciprocal lattice vector $B$.
These two plane-wave states are hybridized in the presence of the
nonzero Fourier component $V_B$, leading to the level repulsion at
the crossing point and yielding the Floquet-Bloch wavefunctions.
Consequently, the quadratic spectrum breaks into loops winding
around in the FBZ.

Fig. \ref{fig:zonefolding} demonstrates an example of zone-folding
of the dispersion corresponding to the Fig 1. (a) in the main text.
Fig. \ref{fig:zonefolding} (a) presents an unfolded dispersion
in the presence of a weak space-time potential. 
The gap opening points are marked in pairs on the unfolded dispersion:
The dashed arrows linking two points represent the non-vanishing Fourier
components $V_B$ of the external potential.
In Fig. \ref{fig:zonefolding} (b), the extended MEBZ representation
is used, i.e., the dispersion is duplicated being shifted by all
the momentum-energy reciprocal lattice vectors. 
The FMEBZ boundary is marked by the solid line.
The Floquet-Bloch dispersion forming two loops after folded into
the FMEBZ as shown in Fig. \ref{fig:zonefolding} (c).
The parts of the spectrum from $2\rightarrow 3$ and from $3^\prime \to 4^\prime$
does not participate in forming these loops, and are omitted.
We can smoothly vary the loop structure in the FMEBZ such that
the winding numbers of each loop become transparent: The red and blue
loops have the winding numbers $(0, 1)$ and $(1, 0)$, respectively.
These two loops actually cross, however, they do not open the gap
due to the lack of non-zero Fourer component $V_B$ with $B$ connecting
two momentum-energy vectors at the crossing point as shown in Fig.
\ref{fig:zonefolding} (a).
Otherwise, we will arrive at the situation shown in Fig 1. (b) in
the main text, where the two loops merge into one  with the winding
number $(1,1)$.

The above example demonstrates the connection between the winding
numbers and the non-vanishing Fourier components of the external
potential. 
The crossing of bands can be protected from splitting by space-time
symmetries of the system.
For example, as the example shown in the main text, the glide
time-reversal $g_t$ protects the spectral double degeneracy.

\section{Construction of the space-time group via group cohomology}
Each space group is constructed from a static Bravais lattice.
Similarly, each space-time group is constructed from a Bravais lattice $M$
constituted of the $D+1$ dimensional space-time mixed discrete
translations, and a magnetic point group in $d$ dimensions
$G_m(D)$ that leaves $M$ invariant.
Here ``magnetic" refers to the reflection with respect to time,
{\it i.e.}, time-reversal operation.
Below we use denote the space-time group by the symbol $ST$. 

\subsection{The general procedure}
The hierarchical classification scheme of the space-time group starts with
the {\it crystal systems}.
Each crystal system is labeled by a set of {\it Bravais lattices}
$\{M\}$ and these lattices share the same magnetic point group symmetry.
Below we use the same symbol $M$ to represent the lattice translation
group associated to the Bravais lattice $M$.
The magnetic point group symmetry of a crystal is often smaller than
the that of the underlying Bravais lattice.
As a result, each crystal system can be further divided into different
{\it geometry} crystal classes (GCC) according to different magnetic
point group symmetries.
Each GCC can be further classified into different arithmetic crystal
classes (ACC) based on a particular Bravais lattice and a particular
magnetic point group.
It is worth noting that, the same Bravais lattice and magnetic point
group can give rise to different ACCs depending on the realizations
of magnetic point group operations.
For each ACC, based on its unique Bravais lattice and magnetic point
group, different space-time groups can be constructed.
Such hierarchy is in parallel to the
space group classification scheme of static lattices.\cite{Prince2004}.

\begin{table}[hptb]
\setlength\extrarowheight{4pt}
\centering
\begin{tabular}{|c|c|c|}
\hline
 Point Group                              &$G_m(2)$            & Generators             \\ \hline
 \multirow{2}{*}{$C_1$}         &$1$                &                        \\ \cline{2-3}
                         &$11'$            &$m_t$                    \\ \hline
 \multirow{3}{*}{$D_1$}         &$m$            &$m_x$                \\ \cline{2-3}
                         &$m1'$            &$m_x, m_t$                \\ \cline{2-3}
                         &$m'$            &$m_xm_t$                \\ \hline
\multirow{3}{*}{$C_2$}        &$2$                &$R_\pi$                  \\ \cline{2-3}
                        &$21'$            &$R_\pi, m_t$              \\ \cline{2-3}
                        &$2'$                &$R_\pi m_t$              \\ \hline
\multirow{4}{*}{$D_2$}               &$mm2$            &$m_x, m_y$                  \\ \cline{2-3}
                        &$mm21'$        &$m_x, m_y, m_t$           \\ \cline{2-3}
                        &$m'm2'$            &$m_xm_t, m_y$          \\ \cline{2-3}
                        &$m'm'2$            &$m_xm_t, m_ym_t$    \\ \hline
\multirow{2}{*}{$C_3$}        &$3$                &$R_{2\pi/3}$                   \\ \cline{2-3}
                                            &$31'$            &$R_{2\pi/3}, m_t$        \\ \hline
\multirow{3}{*}{$D_3$}        &$3m$            &$R_{2\pi/3}, m_x$       \\ \cline{2-3}
                                            &$3m1'$            &$R_{2\pi/3}, m_x, m_t$    \\ \cline{2-3}
                                            &$3m'$            &$R_{2\pi/3}, m_xm_t$    \\ \hline
\multirow{3}{*}{$C_4$}        &$4$                &$R_{\pi/2}$                   \\ \cline{2-3}
                                            &$41'$            &$R_{\pi/2}, m_t$            \\ \cline{2-3}
                                            &$4'$            &$R_{\pi/2}m_t$            \\ \hline
\multirow{4}{*}{$D_4$}        &$4mm$            &$R_{\pi/2}, m_x$               \\ \cline{2-3}
                                            &$4mm1'$        &$R_{\pi/2}, m_x, m_t$    \\ \cline{2-3}
                                            &$4'm'm$            &$R_{\pi/2}m_t, m_x$    \\ \cline{2-3}
                                            &$4m'm'$            &$R_{\pi/2}, m_xm_t$    \\ \hline
\multirow{3}{*}{$C_6$}        &$6$                &$R_{\pi/3}$                   \\ \cline{2-3}
                                            &$61'$            &$R_{\pi/3}, m_t$            \\ \cline{2-3}
                                            &$6'$            &$R_{\pi/3}m_t$            \\ \hline
\multirow{4}{*}{$D_6$}        &$6mm$            &$R_{\pi/3},m_x$               \\ \cline{2-3}
                                            &$6mm1'$        &$R_{\pi/3},m_x, m_t$    \\ \cline{2-3}
                                            &$6'm'm$            &$R_{\pi/3}m_t,m_x$    \\ \cline{2-3}
                        &$6m'm'$            &$R_{\pi/3},m_xm_t$    \\ \hline

\end{tabular}
\caption{The 2D magnetic point groups and their relations to the usual
ten 2D point groups.
The magnetic groups are denoted by using the international notation,
where $'$ means the time reversal operation.
In the third column, the symmetry generators for each magnetic point
group are listed.
The symmetry operation $m_t$ stands for the time reversal, $R_\theta$
represents rotation in the $x$-$y$ plane at the angle of $\theta$,
$m_{x}$ and $m_y$ are the reflections with respect to the $x$- and
$y$-directions, respectively.
The symbols of the magnetic group with $'$
}
\label{tb:mp2D}
\end{table}

\begin{table}[hptb]
\setlength\extrarowheight{5.5 pt}
\centering
\begin{tabular}{|c|c|c|c|c|}
\hline
 Crystal System                                   &Bravais Lattice                    &MP Group                    &ACC                 & $ST(2,1)$        \\ \hline
 Triclinic                                 &Primitive                            &$1, 2'$                           &$2$            &$2$                 \\ \hline
 \multirow{2}{*}{T-Monoclinic}         &Primitive                            &\multirow{2}{*}{$11', 2, 21'$}        &$3$                &$8$            \\ \cline{2-2} \cline{4-5}
                             &Centered                        &                            &$3$                &$5$                \\ \hline
 \multirow{2}{*}{R-Monoclinic}         &Primitive                            &\multirow{2}{*}{$m, m', m'm2'$}    &$3$                &$8$            \\ \cline{2-2} \cline{4-5}
                             &Centered                        &                            &$3$                &$5$                \\ \hline
\multirow{5}{*}{Orthorhombic}        &Primitive        &\multirow{5}{*}{\pbox{20cm}{$mm2, m'm'2$\\$mm21', m1'$}}    &$4$                &$68$          \\ \cline{2-2} \cline{4-5}
                            &T-Base-Centered                    &                            &$4$                &$15$              \\ \cline{2-2} \cline{4-5}
                            &R-Base-Centered                    &                            &$5$                &$22$              \\ \cline{2-2} \cline{4-5}
                            &Face-Centered                    &                            &$4$                &$7$              \\ \cline{2-2} \cline{4-5}
                            &Body-Centered                    &                            &$4$                &$15$            \\ \hline
\multirow{2}{*}{Tetragonal}                 &Primitive            &\multirow{2}{*}{\pbox{20cm}{$4,41', 4'$\\$4mm, 4mm1'$\\$4'm'm, 4m'm' $}}&$8$    &$49$              \\ \cline{2-2} \cline{4-5}
                            &Body-Centered    &                                                           &$8$    &$19$              \\ \hline
\multirow{2}{*}{Trigonal}            &Primitive            &\multirow{2}{*}{\pbox{20cm}{$3, 6', 3m$ \\$3m' , 6'm'm$}}               &$8$    &$18$               \\ \cline{2-2} \cline{4-5}
                                                &Rhombohedral    &                                                           &$5$    &$7$                \\ \hline
Hexagonal                          &Primitive            & \pbox{3cm}{$6, 61', 31'$\\$6mm, 6m'm'$\\$6mm1', 3m1'$}                &$8$    &$27$            \\ \hline
\end{tabular}
\caption{Classification of the 2+1D space-time groups.
There are 7 space-time crystal systems, 14 space-time
Bravais lattices (The primitive trigonal lattice is
the same as the primitive hexagonal lattice).
The 31 magnetic point groups are uniquely assigned to
the 7 crystal systems, as listed in the third column.
In the 4th and 5th columns, the numbers of the ACC
and the space-time groups for each Bravais lattice
are listed.
}
\label{tb:sptgrps2D}
\end{table}

\subsection{The method of group cohomology}
Given an ACC labeled by $M$ and $G_m$, where $M$ is a Bravais lattice and
$G_m$ is a magnetic point group symmetry, all its space-time groups can
be classified via the method of group cohomology theory, similar to the
space group classifications \cite{Hiller1986}.
A space-time group $ST$ is the group extension of the lattice
translation group of $M$ by the magnetic point group $G_m$,
described by the following exact sequence.
\bea
1\rightarrow M \rightarrow ST \rightarrow G_m\rightarrow 1,
\eea
where $\rightarrow$ means a mapping.
For two consecutive mappings, the image of the first mapping is the
same as the kernel of the second one.
Such a group extension can be constructed by associating each magnetic
point group element $g$ in $G_m$ by a $D+1$ dimensional fractional
translation $c(g)$ not belonging to the lattice translation group.
$c(g)$ can be viewed as a mapping from $G_m$ to $T_{D+1}/M$
where $T_{D+1}$ is the continuous translation group in $D+1$ dimensions.
$c(g)$ satisfies
\bea
c(1)=0, \ \ c(g_1g_2)=c(g_1)+g_1c(g_2).
\label{eq:cocycle}
\eea
We denote $(c(g),g)$ as the combined operation of first applying $g$ followed
by $c(g)$, then the set of these elements form a group following the rule
of product as
\bea
(c(g_1), g_1) (c(g_2),g_2)=(c(g_1g_2), g_1g_2).
\eea

To classify all the space-time groups for the same ACC, a key observation is
that all the different ways of mapping themselves form an Abelian group.
Given two distinct mapping $c$ and $d$, we define their product $c\cdot d$
as
\bea
c\cdot d (g)=c(g)+d(g),
\eea
which satisfies Eq. \ref{eq:cocycle} as well.
This group is denoted as $Z^1(G_m, T_{D+1}/M)$.
However, not all the elements in $Z^1(G_m, T_{D+1}/M)$ correspond to distinct
types of space-time groups.
Without specifying the equivalence relations, $Z^1(G_m, T_{D+1}/M)$
is not finite.
An obvious condition is that space-time groups that related by shifting the
origin are of the same type.
For example, the map of
\be
c_u(g)=gu-u,
\label{eq:coboundary}
\ee
simply shifts the magnetic point group operation origin to $u \in T_{D+1}/M$,
which should be identified with the trivial map $c(g)=0$.
The maps with the structure in Eq. \ref{eq:coboundary} form a group
$B^1(G_m, T_{D+1}/M)$.
The first cohomology group of $G_m$ with coefficients in $T_{D+1}/M$
is defined as the quotient group of
\bea
H^1(G_m, T_{D+1}/M)=Z^1(G_m, T_{D+1}/M)/B^1(G_m, T_{D+1}/M),
\nn \\
\eea
Each element of $H^1(G_m, T_{D+1}/M)$ corresponds to one space-time group
of a particular ACC characterized by the magnetic point group $G_m$ and
the Bravais lattice $M$.

The identity of $H^1(G_m, T_{D+1}/M)$ is the trivial map that $c(g)=0$
for all $g$ in the magnetic point group $G_m$.
The corresponding space-time group is the semi-direct product of the
lattice translation group $M$ and the magnetic point group $G_m$, which
is called the ``symmorphic" space-time group.
By definition, each ACC only contains one symmorphic space-time group.
Other elements in $H^1(G_m, T/M)$ correspond to ``nonsymmorphic" space-time
groups.
For a crystal with ``nonsymmorphic" space-time group symmetries, it is
typically not invariant under the magnetic point group operations in $G_m$.

However, not all the elements of the cohomology group $H^1(G_m, T_{D+1}/M)$
lead to distinct space-time groups either.
For example, two elements in $H^1(G_m, T_{D+1}/M)$ related by a global
rotation should be identified.
Hence, we invoke the second equivalence relation that all elements
in $H^1(G_m, T_{D+1}/M)$ related by linear transformations $\rho$ are
identified, where $\rho$ leaves \textit{all} the magnetic point groups
of the crystal system unchanged.

\section{Classification of the 2+1D space-time group}

\begin{figure}
\includegraphics[height=0.35\columnwidth,width=0.4\columnwidth]{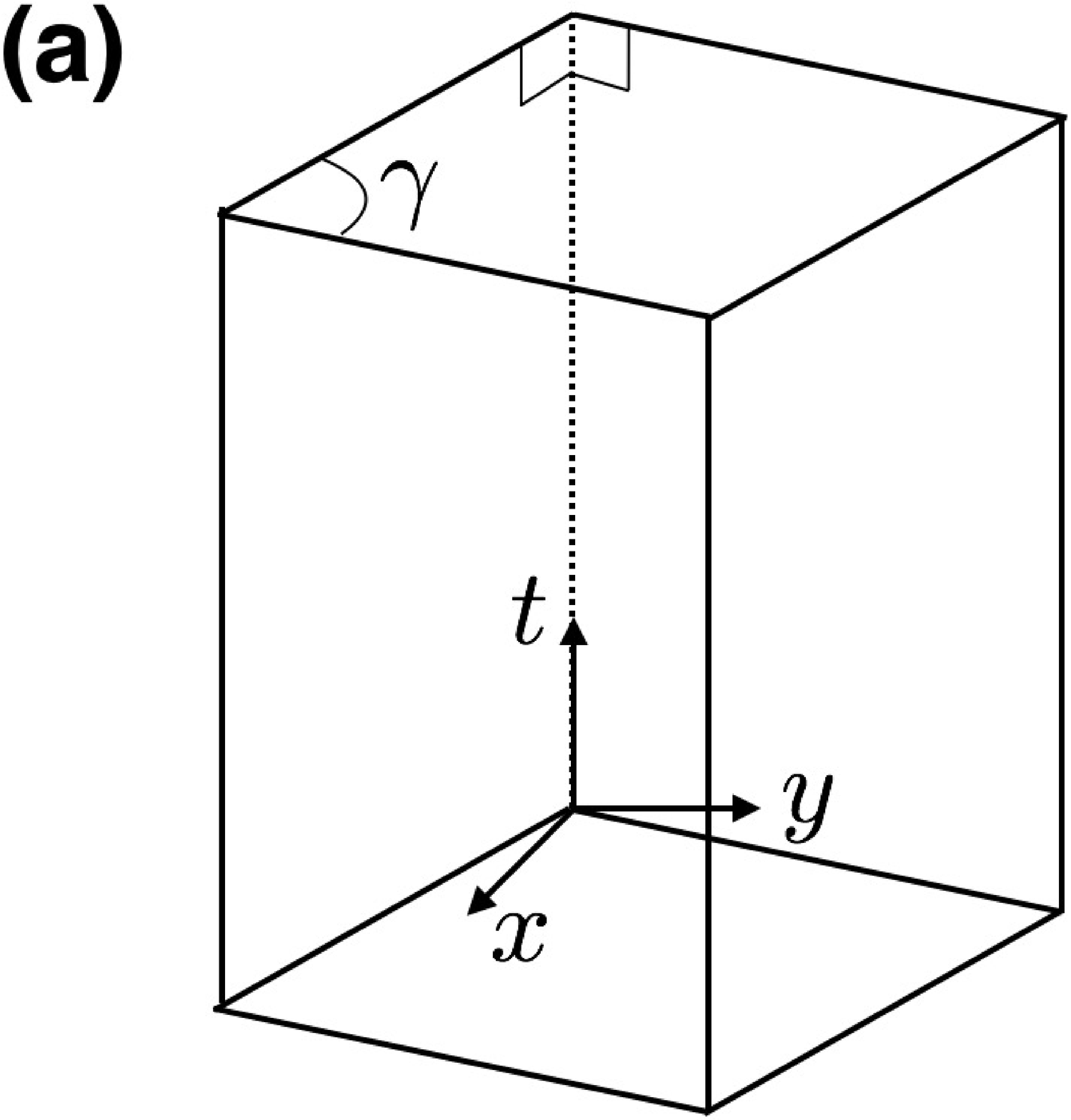}
\includegraphics[height=0.35\columnwidth,width=0.35\columnwidth]{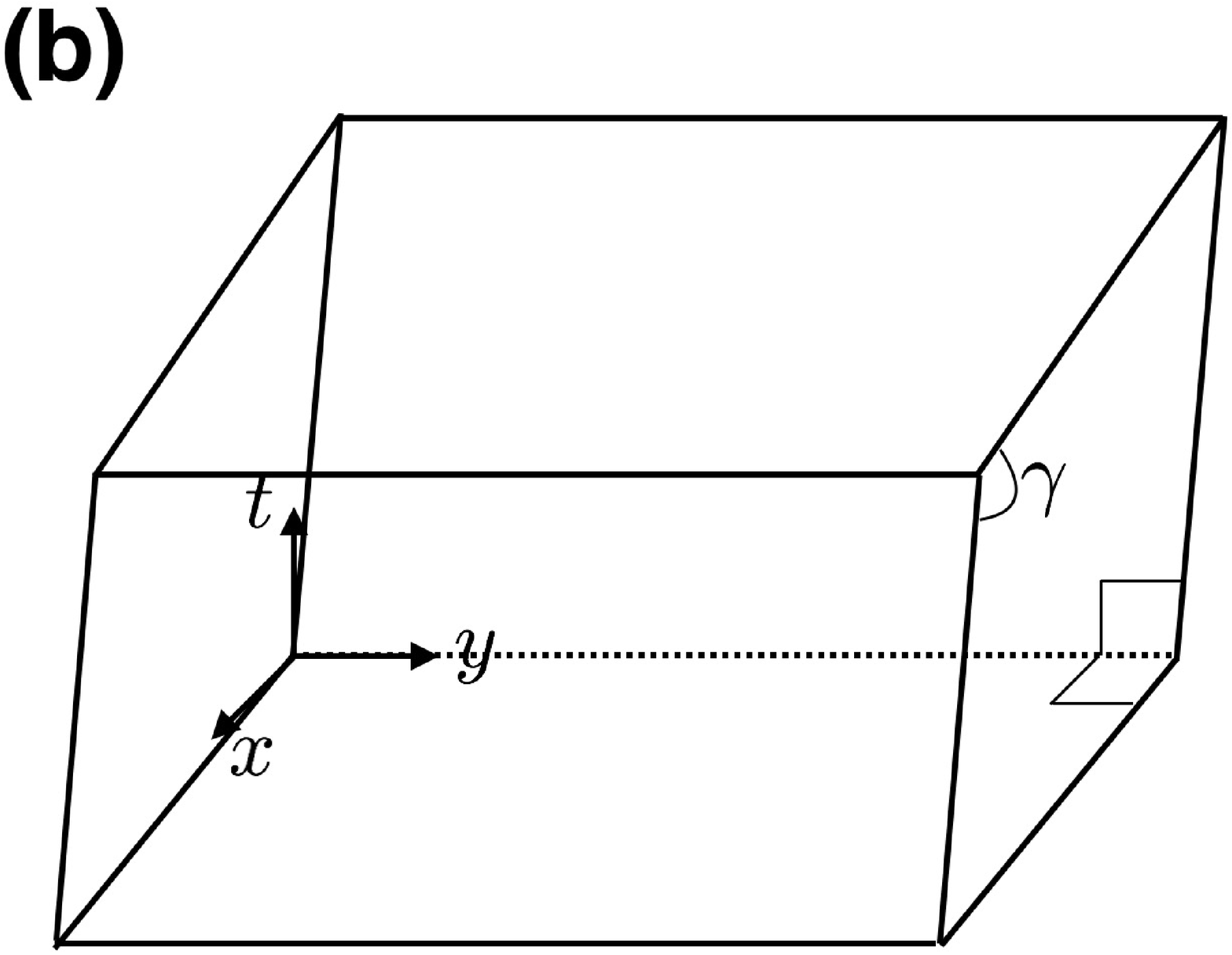}
\caption{The different types of monoclinic space-time crystal systems.
(a) The $t$-monoclinic crystal system.
The $c$-axis is along the temporal direction and perpendicular to
the $xy$-plane.
(b) The $r$-monoclinic crystal system.
The $c$-axis is along one spatial direction, say, the $y$-direction,
and is perpendicular to the $xt$-plane.
The maximal magnetic point groups for (a) and (b)
are $21'$ and $m'm2$, respectively.
}
\label{fig:mono}
\end{figure}

\begin{figure}
\includegraphics[height=0.35\columnwidth, width=0.35\columnwidth]
{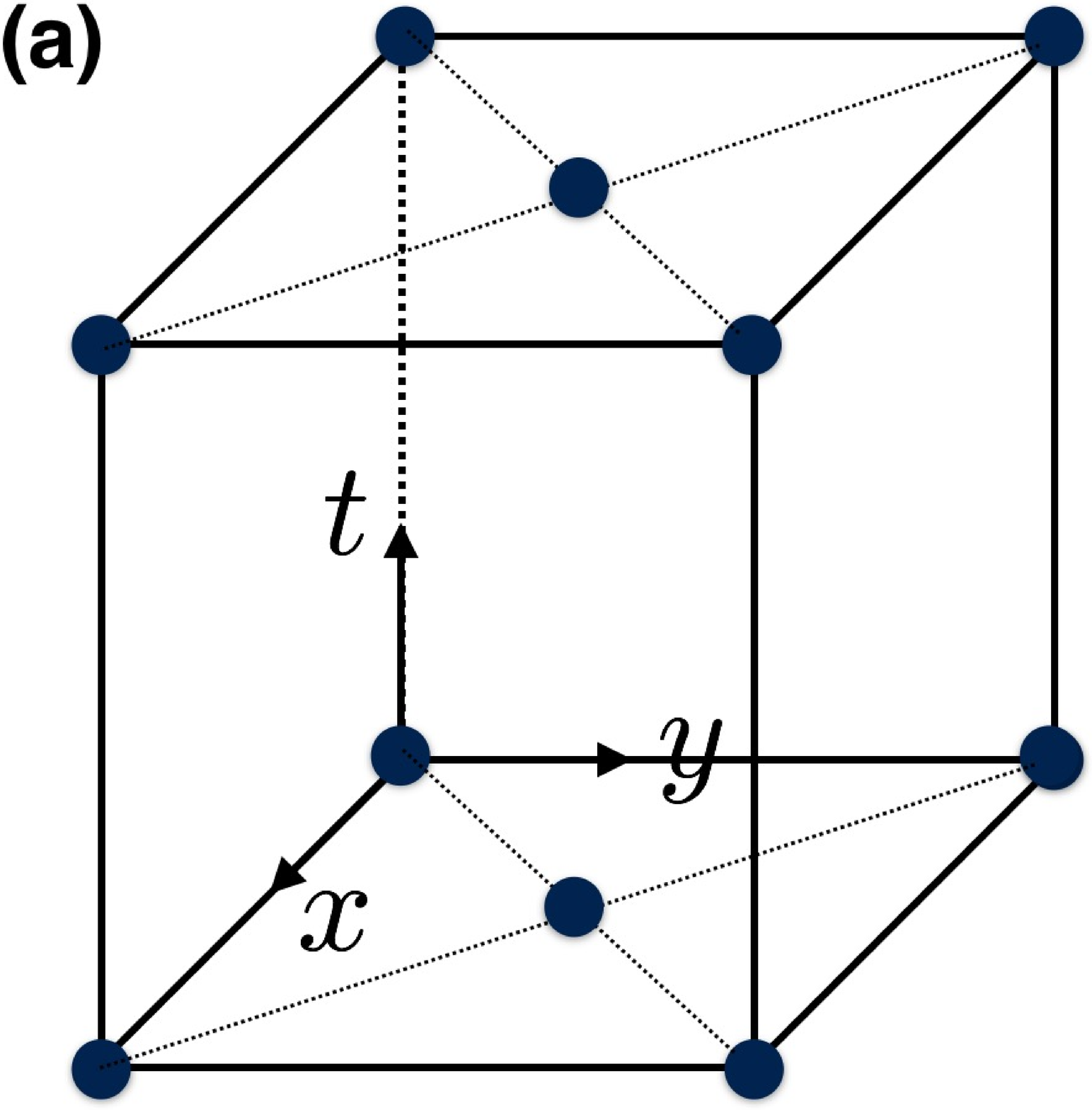}
\includegraphics[height=0.35\columnwidth, width=0.35\columnwidth]
{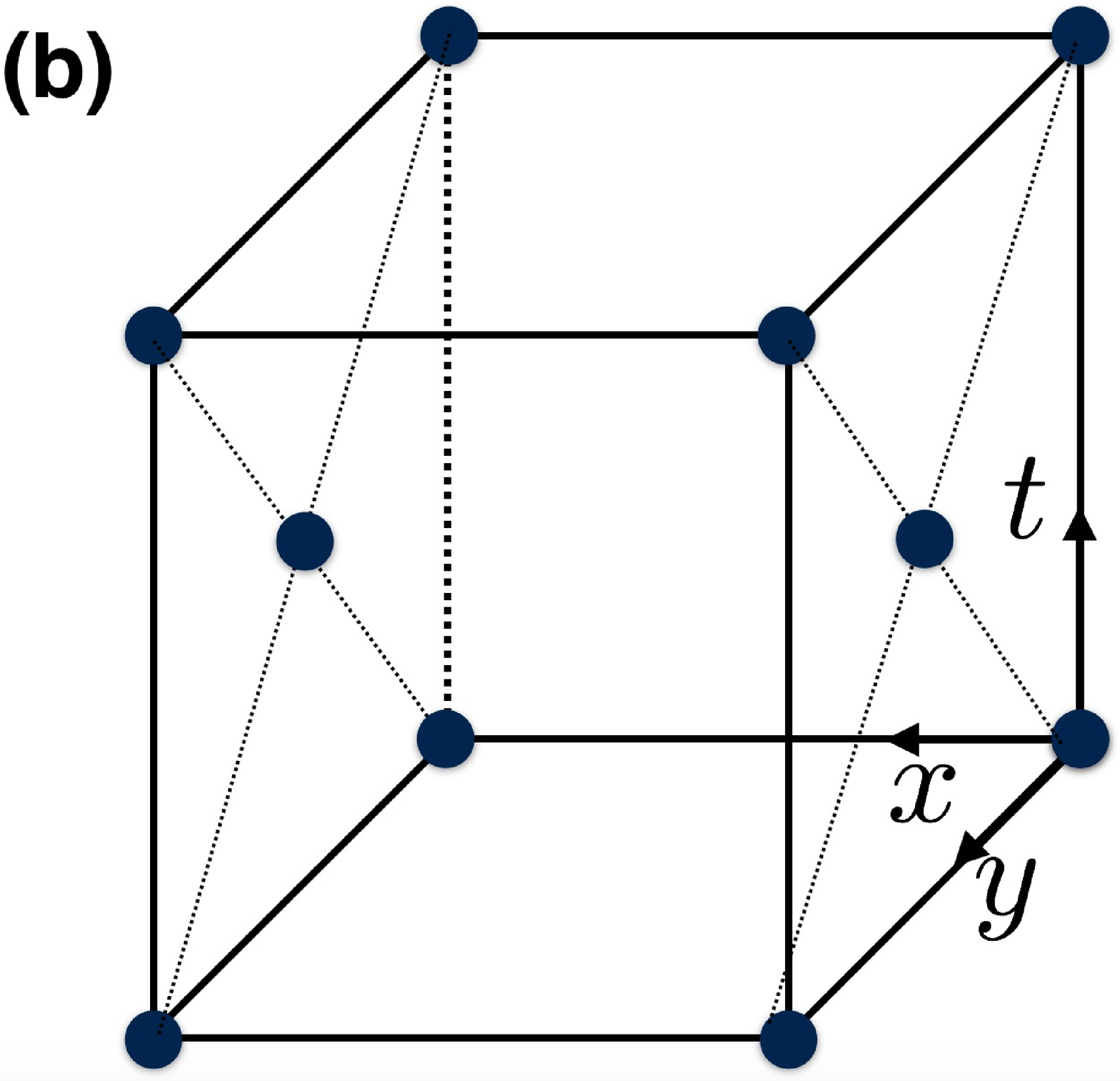}
\caption{The two types of base-centered Bravais lattices in the
space-time orthorhombic crystal system.
(a) The $t$-base-centered lattice, and (b) the $r$-base-centered
lattice.
The centered bases lie in the $xy$ and $xt$ planes, respectively.
}
\label{fig:base_orthg}
\end{figure}

In the main text, we list the classification of the space-time groups
in 1+1 dimensions as a proof of concept.
In this section, we focus on the case of 2+1 dimensions.
The case of 3+1 dimensions is left for future study.

In 2 dimensions, there exist 10 crystallographic point groups, which give
rise to 31 magnetic point groups as listed in Table. \ref{tb:mp2D}.
By combining the magnetic point groups with the discrete translation
symmetries in $T_{2+1}$, we obtain 7 space-time crystal systems, 14
Bravais lattices, and 275 space-time groups.
The relation among space-time crystal systems, Bravais lattices,
magnetic point groups, and space-time crystals is summarized in
Table \ref{tb:sptgrps2D}.

We adopt the terminology mostly from crystallography for 3D static
systems \cite{Aroyo2016}.
The detailed structure of each space-time group structure will be
published somewhere else.
The difference between the 2+1D space-time group and the 3D static space
groups is outlined below.
The main point is the specialty of the temporal direction: The
time reversal operation is anti-unitary, and space-time mixed
rotations are not allowed in the space-time group.
The second factor significantly affects the classification.
As shown in the first column of Table. \ref{tb:sptgrps2D}, the
difference already appears on the level of crystal systems.
Since the temporal and spatial directions are non-equivalent,
the cubic crystal system is absent in 2+1 D, and in addition,
there are two kinds of monoclinic crystal systems, $r$-monoclinic
and $t$-monoclinic, depending on whether the $c$-axis is along
the temporal or spacial directions.

The primitive unit cells for both $r$- and $t$-monoclinic space-time
crystal systems in Fig. \ref{fig:mono}.
The maximal magnetic point groups of these two crystal systems
are nonequivalent.
In the first case, it is $21'$ generated by the spatial rotation
$R_\pi$ and time reversal $m_t$, while it is $m'm2$ in the second
case generated by $m_ym_t, m_x$.
Furthermore, within the orthorhombic space-time crystal system,
there are two different base-centered Bravais lattices illustrated
in Fig. \ref{fig:base_orthg}, depending on the centered bases lying
in the $xy$ and $x(y)t$-planes, respectively.
The former is denoted as $r$-base-centered lattice, and the latter
is $t$-base-centered lattice.
As a result, there exist 5 Bravais lattices in the orthorhombic space-time
crystal system rather than 4 in its 3D static counterpart.

In these 2+1D space-time crystal systems, the number of the space-time groups
within each specific ACC is also in general different from its 3D
static counterpart.
On the other hand, when it comes to the tetragonal, the trigonal
and the hexagonal crystal systems, there is a one-to-one
correspondence between the 2+1 D space-time groups and the 3D
space groups, since the rotation plane must be purely spatial.

\section{Protected degeneracies due to non-commutative symmetry operators}
In the section, we discuss the protected band structure degeneracy
occurring at high symmetric points in MEBZ due to the non-symmorphic
nature of the space-time group symmetry.

A generic group element $g$ in the space-time group takes the form,
\be
g=T_\mathbf{r}(\mathbf{u})T_t(\tau) R m_t^s
\ee 
where $T_\mathbf{r}(\mathbf{u})$ is the spatial translation, $T_t(\tau)$
is the temporal translation; $R$ is a point group operator
acting only on the spatial dimensions; $m_t$ is the anti-unitary
time-reversal operation;
$s=1$ or 0 determines whether $g$ includes time-reversal and is anti-unitary.

Consider two operations $g_1$ and $g_2$ in the little group of a
high symmetric point $\kappa=(\mathbf{k},\omega)$.
The degeneracy condition at $\kappa$ can be obtained by commuting
these two operators.
After some algebra, we arrive at 
\bea
g_1g_2&=&T_\mathbf{r}(\tilde{\mathbf{u}})T_t(\tilde{t})g_2g_1\tilde{R}
\eea
with
\bea
\tilde{R}&=&R_1^{-1}R_2^{-1}R_1R_2, \nonumber \\
\mathbf{\tilde{u}}&=&(\mathbf{I}-R_2)\mathbf{u_1}
-(\mathbf{I}-R_1)\mathbf{u_2},  \nn \\
\tilde{t}&=&2 s_2 t_1-2 s_1 t_2.
\label{SM:degeneracy}
\eea
We assume $R_1$ and $R_2$ commute, i.e., $\tilde{R}=\mathbf{I}$, then
their operations on Floquet-Bloch wavefunctions with $\kappa$ satisfy
\bea
M_{g_1}M_{g_2}=
e^{i \mathbf{k}\cdot\tilde{\mathbf{u}}-i \omega \tilde{t}}M_{g_2} M_{g_1}.
\label{eq:commute}
\eea
Depending on whether $g_1$ and $g_2$ are unitary or anti-unitary,
there are three different cases.

First, if neither of $g_1$ and $g_2$ flips the direction of time,
i.e., both $M_{g_1}$ and $M_{g_2}$ are unitary, 
then $\tilde{t}=0$ and the phase factor in Eq. (\ref{eq:commute})
is independent of $\omega$.
This is the situation which has been studied in the main text.

Second, one of the $g$ operators, without loss of generality,
say, $g_1$, flips the direction of time and the other does not.
In this case, $M_{g_1}$ is anti-unitary while $M_{g_2}$ is unitary. 
Now $\tilde{t}=2t_1$, and the prefactor in Eq. \ref{eq:commute} 
does have frequency dependence.
However, due to the involving of the anti-unitary operator, the degeneracy
condition is more subtle than the previous case and shall
be discussed more carefully.
Consider a Floquet-Bloch state $\psi_\kappa$ as an eigenstate of
the unitary operator $g_2$,
\be
M_{g_2} \psi_\kappa=e^{i \mathbf{k}\cdot \mathbf{u_2}-i \omega t_2}
e^{i\theta} \psi_\kappa,
\ee
in which we explicitly separate the phase dependence on $\kappa$
, and $\theta$ \textit{only} depends on the point group
operation $R_1$. 
Based on Eq. \ref{eq:commute}, one can show that $M_{g_1} \psi_\kappa$
is also an eigenstate of $g_2$ with, in principle, a different eigenvalue,
\be
M_{g_2} M_{g_1} \psi_\kappa=e^{-i\mathbf{k}\cdot (\mathbf{u_2}+\tilde{\mathbf{u}})-i \omega t_2}e^{-i\theta} M_{g_1}  \psi_\kappa.
\ee
Therefore, if the two phases do not equal, i.e.
$e^{i \mathbf{k}\cdot (2\mathbf{u}_2+\tilde{\mathbf{u}})+2i\theta}\neq1$,
the space-time symmetry-protected degeneracy occurs.
Similar to the previous case, the degeneracy condition does not depend
on the frequency component of $\kappa$.
The last case is when both $g_1$ and $g_2$ flip the time-direction,
i.e., both $M_{g_1}$ and $M_{g_2}$ are anti-unitary.
This can be reduced to the 2nd case by defining $g_1^\prime=g_1g_2$,
whose $M_{g_1^\prime}$ is unitary again. 
Then $g_1^\prime$ and $g_2$ satisfy
\bea
g_1^\prime g_2=T_{\mathbf r} (\mathbf{u}) T_t (\tau) g_2 g_1^\prime,
\eea
where $\mathbf{u}$ and $\tau$ are defined according in
Eq. \ref{SM:degeneracy}.

In short, the degeneracy condition arising from the space-time symmetries
considered here does not depend on the frequency.
This is expected since one can always shift the frequency of the spectrum
by adding a constant to the time-dependent Hamiltonian.
In principle, one could also study the degeneracy condition resulting
from the interplay between the space-time symmetry and the chiral symmetry.


\end{document}